\def\year{2025}\relax
\documentclass[letterpaper]{article} \usepackage{aaai25}  \usepackage{times}  \usepackage{helvet}  \usepackage{courier}  \usepackage[hyphens]{url}  \usepackage{graphicx}    \usepackage{natbib}  \usepackage{caption}    

\usepackage{bm}
\usepackage{amsmath}
\usepackage{xfrac}
\usepackage{amsfonts}
\usepackage{xspace}
\usepackage[capitalise,nameinlink]{cleveref}
\usepackage{subfig}
\usepackage[export]{adjustbox}

\newcommand{\answerYes}[1]{\textcolor{blue}{#1}} 
\newcommand{\answerNo}[1]{\textcolor{teal}{#1}} 
\newcommand{\answerNA}[1]{\textcolor{gray}{#1}}

\usepackage{booktabs}
\usepackage{multirow}
\usepackage{comment}
\usepackage{tabularx}
\usepackage{rotating}
\usepackage{makecell}
\usepackage{colortbl}

\usepackage{xparse}
\usepackage{scalerel,graphicx}
\NewDocumentCommand\emojiAUS{}{\scalerel*{\includegraphics{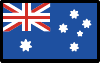}}{X}\xspace}
\NewDocumentCommand\emojiUS{}{\scalerel*{\includegraphics{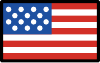}}{X}\xspace}  
\NewDocumentCommand\emojiRAINBOW{}{\scalerel*{\includegraphics{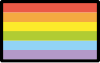}}{X}\xspace}
\NewDocumentCommand\emojiCOFFEE{}{\scalerel*{\includegraphics{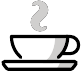}}{X}\xspace}

\usepackage{microtype}

\usepackage[colorinlistoftodos,textsize=tiny]{todonotes}
\definecolor{navy}{rgb}{0.1, 0.1, 0.8}
\definecolor{gray}{rgb}{0.4, 0.4, 0.4}
\definecolor{lightgray}{rgb}{0.8, 0.8, 0.8}
\definecolor{olive}{rgb}{0.1, 0.5, 0.1}
\definecolor{ruby}{rgb}{0.8, 0.1, 0.3}

\usepackage{soul}
\newcommand{\eat}[1]{}

\newcommand{\revR}[1]{{#1}}

\newcommand{\mar}[1]{}

\newcommand{\titlename}{Practical Guidelines for Ideology Detection Pipelines and Psychosocial Applications}

\title{\titlename}
\author{
    Rohit Ram\textsuperscript{\rm 1,2},
    Emma Thomas\textsuperscript{\rm 3},
    David Kernot\textsuperscript{\rm 4},
    Marian-Andrei Rizoiu\textsuperscript{\rm 2}
}
\affiliations{
    \textsuperscript{\rm 1} Thaum\\
    \textsuperscript{\rm 2} University of Technology Sydney\\
    \textsuperscript{\rm 3} Flinders University\\
    \textsuperscript{\rm 4} Defence Science and Technology Group\\
    rohit@thaum.io,marian-andrei.rizoiu@uts.edu.au,\\
    emma.thomas@flinders.edu.au,david.kernot@defence.gov.au
}

\makeatletter

\def\somemagiccommand#1{\let\tc@w\@empty
\protected@edef\tmp{\noexpand\tc@a#1\relax}\expandafter\tc@uc@\tmp}

\def\tc@a{\futurelet\tmp\tc@aa}

\def\tc@aa{\ifcat a\noexpand\tmp\expandafter\tc@ab
\else\expandafter\tc@ac\fi}

\def\tc@ab#1{\edef\tc@w{\tc@w#1}\tc@a}

\def\tc@ac{\csname tc@@\tc@w\endcsname\expandafter\tc@uc\tc@w
\let\tc@w\@empty
\ifx\tmp\@sptoken\let\next\tc@sp
\else\ifx\tmp\relax\let\next\relax
\else\let\next\tc@nxt
\fi\fi\next}

\def\tc@sp#1{ \tc@a#1}
\def\tc@nxt#1{#1\tc@a}

\def\tc@uc#1{\uppercase{#1}}
\def\tc@uc@#1#2{\uppercase{#1#2}}

\let\tc@@the\@gobbletwo
\let\tc@@and\@gobbletwo

\makeatother

\newcommand{\qanda}{\#QandA\xspace}
\newcommand{\ausvotes}{\#Ausvotes\xspace}
\newcommand{\socialsense}{\#Socialsense\xspace}
\newcommand{\parler}[1]{Parler\xspace}
\newcommand{\riot}[1]{Riot\xspace}

\newcommand{\gr}[1]{{\color{gray}{#1}}}

\newcommand{\sh}[1]{\noindent \textbf{\somemagiccommand{#1}}}

\newcommand{\hashtags}{\textsc{Hashtags}\xspace}
\newcommand{\communityparticipation}{\textsc{Community Participation}\xspace}
\newcommand{\media}{\textsc{Media}\xspace}
\newcommand{\politicianHOP}{\textsc{Politician Endorsers}\xspace}
\newcommand{\politicalParties}{\textsc{Party Followers}\xspace}
\newcommand{\urlLR}{\textsc{Left-Right MPP}\xspace}
\newcommand{\urlFR}{\textsc{Far-Right MPP}\xspace}
\newcommand{\mbfc}{\textsc{MBFC MPP}\xspace}

\begin{document}

\maketitle

\begin{abstract}
    Online ideology detection is crucial for downstream tasks, like countering ideologically motivated violent extremism and modeling opinion dynamics.
    However, two significant issues arise in practitioners' deployment.
    Firstly, gold-standard training data is prohibitively labor-intensive to collect and has limited reusability beyond its collection context (i.e., time, location, and platform).
    Secondly, to circumvent expense, researchers employ ideological signals (such as hashtags shared). 
    Unfortunately, these signals' annotation requirements and context transferability are largely unknown, and the bias they induce remains unquantified.
    This study provides guidelines for practitioners requiring real-time detection of left, right, and extreme ideologies in large-scale online settings.
    We propose a framework for pipeline constructions, describing ideology signals by their associated labor and context transferability.
    We evaluate many constructions, quantifying the bias associated with signals and describing a pipeline that outperforms state-of-the-art methods ($0.95$ AUC ROC).
    We showcase the capabilities of our pipeline on five datasets containing more than 1.12 million users.
    We set out to investigate whether the findings in the psychosocial literature, developed for the offline environment, apply to the online setting.
    We evaluate at scale several psychosocial hypotheses that delineate ideologies concerning morality, grievance, nationalism, and dichotomous thinking.
    We find that right-wing ideologies use more vice-moral language, have more grievance-filled language, exhibit increased black-and-white thinking patterns, and have a greater association with national flags.
    This research empowers practitioners with guidelines for ideology detection, and case studies for its application, fostering a safer and better understood digital landscape.
\end{abstract}

\begin{links}
\link{Code}{github.com/behavioral-ds/ideology_prediction}
\end{links}

 \section{Introduction}
\label{ide:sec:introduction}

Ideologies are the collection of beliefs and opinions about the ideal arrangement of society \cite{cohrs2012ideological}. 
Tracking extreme ideologies is particularly important in detecting extreme voices that can spread harmful and false information, leading to dangerous and even deadly outcomes.
Ideology is canonically (and inexactly) projected onto a left-right spectrum, where the left is associated with equality and reform, and the right is associated with authority and tradition. 
There has been a recent increase in fringe and extreme-leaning worldviews, including the far-right -- a prominent archetype of extreme ideologies associated with ultranationalism and opposition to multiculturalism.
Worryingly, this has increased Ideologically Motivated Violent Extremism (IMVE)~\citep{SAFETY2022} -- a term coined to encompass religious, political and nationalist extremism.
Ideology detection is a lead indicator for IMVE, fortifying individual and collective security. 
It facilitates understanding these ideological groups' values and beliefs, which helps design interventions, build political bridges and tackle radicalization.

Radicalization can occur in a matter of weeks \cite{McCauley2008,Booth2024}, both offline (face-to-face) and online (forums and social media platforms).
To combat this, practitioners -- such as law enforcement and national security agencies -- need practical, real-time ideology detection tools that minimize human effort and can be applied across diverse contexts. 
Despite the significant existing literature, practical and effective detection guidelines remain scarce.
This study establishes a framework for ideology detection pipelines, examining diverse constructions and demonstrating practical implementations using off-the-shelf components. 
Our first aim is to identify practical pipelines that reduce annotation efforts while maintaining transferability across different contexts.
Our second aim is to validate insights into the psychosocial asymmetries of ideologies.
We leverage five large datasets, totaling 1.12 million profiles, and test several hypotheses from the psychosocial literature at scale, mainly developed in offline laboratory setups.
We answer two specific research questions.

The first question involves \emph{ideological proxies} -- measurable user behavior signals correlating with ``true'' ideology -- that minimize annotation labor and are transferable across contexts.
We define a \emph{context} as the tuple (topic, time, geography, platform).
Prior works rely on various sources of ideological knowledge, including manually labeling users, labeling ideological proxies, and detecting group behavior differences. 
However, these approaches have limitations: the former two require extensive expert labeling -- an expensive resource -- and often fail to transfer across contexts.
The latter often lacks robustness. Of the three, ideological proxies are the most common approach to reducing labor; however, they vary in reusability.
See \cref{ide:sec:prior-work} for a complete discussion.

Furthermore, few users partake in direct ideological activity, and some actively avoid disclosure.
Consequently, many proxies reveal only the vocal subset of users, biasing downstream analyses \cite{alkiek2022classification,cohen2013classifying}.
Despite this, prior works commonly use proxies as ground truth \cite{darwish2020unsupervised, rashed2021embeddings, xiao2020timme} without quantifying the bias this entices. 
Our first research question is:
\textbf{Which ideological proxies minimize annotation labor, maximize context transferability, and reduce bias?}

The second question involves the psychosocial asymmetries of ideologies.
Understanding the values and beliefs of ideological groups is instrumental in modeling their polarization and user radicalization.
Ideological asymmetry studies are abundant in relevant disciplines \cite{tomkins1963left,jost2017asymmetries}; often shown via offline surveys.
For example, moral values delineate left-from-right ideologies \cite{graham2009liberals}; and grievance/grudge language delineate moderate-from-extreme ideologies \cite{Stankov2021, van2021grievance}.
Many of these hypotheses were developed with offline populations, and there is limited evidence for online populations.
We know online and offline populations differ demographically \citep{auxier2021social}, but we do not understand their psychosocial differences.
We ask \textbf{can we build psychosocial profiles of ideological groups and employ them to evaluate hypotheses related to the psychosocial traits of these groups?}

\sh{Solution outline.}
First, we evaluate ideological proxies. 
We make the widely adopted assumption that homophily -- the tendency of similar individuals to associate -- propagates ideology.
We build a framework to construct ideology detection pipelines.
We qualitatively evaluate proxies, by their minimization of labelling efforts and how readily they transfer to new contexts, and quantitatively evaluate proxies on their prediction of human-annotated ground truth.
We show that a pipeline constructed through a proxy based on media consumption and a lens based on text, is both qualitatively advantageous and quantitatively performant.
Second, we use a pipeline to test hypotheses of the psychosocial asymmetries of ideology at scale.
\begin{figure}[tbp]
    \newcommand\myheight{0.29}
    \centering
\includegraphics[width=\columnwidth]{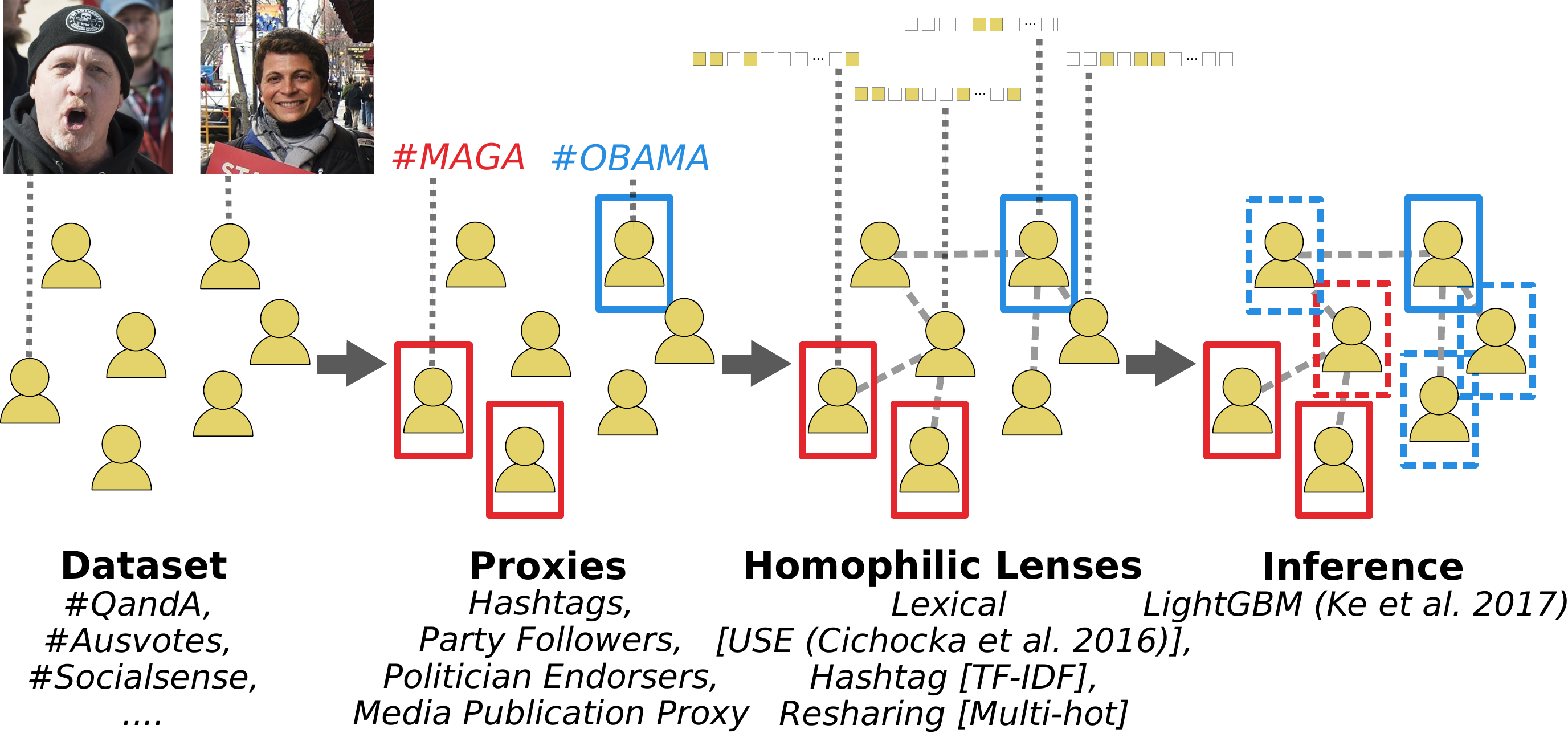}
\caption{    
        \textbf{The schema conceptualizes the four components of the pipeline}; 
        (1) the datasets contain information about users (two examples are shown), 
        (2) the ideological proxies assign labels on some of the users based on external information (here \#MAGA indicates right-leaning, while \#OBAMA indicates left-leaning), 
        (3) the homophilic lenses build numeric descriptions for user and a way to measure their similarity, and 
        (4) inference architecture predicts the likely labels of all other users in the dataset.
    }
    \label{ide:fig:schema}
\end{figure} 
We address the first question in \cref{ide:sec:methodology}.
We introduce our pipeline framework, consisting of four components: dataset, ideological proxy, homophilic lens, and inference architecture. 
We use five \emph{social media datasets}, collected from three platforms (Parler, Facebook and Twitter), containing 1.12 million users, and spanning social domains such as TV shows, elections, climate change, antivaccination and the January 6th US Capitol Insurrection.
We frame the problem as user classification: the left-right detection as ternary classification (left, right, and neutral), and the far-right detection as binary classification.
We limit our scope to Anglo-centric, English-speaking contexts with a dominant uniaxial political spectrum\footnote{
This is not to discount the need for ideology detection in other regions, like the Global South. 
Nor to suggest that a uniaxial spectrum is sufficient to encompass the complex politics of global communities.
See \cref{ide:sec:Conclusion} for a discussion.
}.
We explore four left-right and two far-right \emph{ideology proxies}, leveraging behaviors such as posting politically-charged hashtags, following political parties, endorsing politicians, and sharing media websites. 
We build three \emph{homophilic lenses} based on language, endorsements, and topics.
We use the ideology proxies and homophily lenses to build ideology pipelines with an \emph{off-the-shelf classifier}.
See \cref{ide:fig:schema}.

In \cref{ide:sec:results} we evaluate the performance of ideology detection pipelines.
We construct gold-standard benchmarks for left-right and far-right classification via human annotation and use them to evaluate bias introduced by ideological proxies.
Furthermore, we assess various combinations of ideology proxy and homophilic lens to observe interaction effects and find the best performing combination.
Finally, we compare this pipeline to state-of-the-art baselines: TIMME \cite{xiao2020timme}, UUS \cite{darwish2020unsupervised}, and UUS+ \cite{samih2021few} and achieve the best area-under-the-receiver-operating-curve (AUC ROC) of $0.95$, an improvement of $6.7\%$ over the next best, TIMME.

We address the second question in \cref{ide:sec:psychosocial-results}.
We evaluate psychosocial hypotheses relating to morality, grievance, nationalism, and dichotomous thinking.
For \emph{morality}, we evaluate the seminal Moral Foundations Theory \citep{graham2009liberals} hypotheses, operationalized via FrameAxis \citep{kwak2021frameaxis} (see \cref{ide:sec:preliminaries}). 
In its two subsets of hypotheses, individualizing and binding; we find relatively more support for the prior.
However, only $46\%$ of hypotheses are supported overall. 
As alternative hypotheses, we find that the right uses the language of vice more than the left, with statistical significance.
For \emph{grievance}, following literature that theorized that grudges and grievances are requirements for radicalization~\citep{Stankov2021}, we find large-scale proof that the far-right uses grievance language more than moderates.
We operationalize via the Grievance Dictionary threat-assessment tool \citep{van2021grievance} (see \cref{ide:sec:preliminaries}).
For \emph{nationalism}, we show that the right exhibit nationalism via flag emojis, adding validity to our inferred grouping.
Finally, for \emph{dichotomous thinking}, we apply a dictionary-based approach, showing that the far-right, followed by the right, exhibits more black-and-white thinking (supporting prior work).

The main contributions of this work are as follows:
\begin{itemize}
    \item An ideology detection pipeline applicable in large-scale online setups, that minimizes labor requirements and improves transferability to multiple contexts. \item The most comprehensive discussion and analysis of ideological proxies (to our knowledge); quantifying their bias independently and jointly with homophilic lenses.
    One construction outperforms state-of-the-art methods.
    \item Evaluation of psychosocial hypotheses concerning ideologies in a large-scale online setting. \end{itemize}

\sh{Glossary.}
For readability, we collocate and define terms here.
\textit{Ideological Proxy}: measurable user behaviors correlating with ideology (e.g., emitting hashtags, following ideological users, sharing ideological media, etc.).
\textit{Homophilic Lens}: a representation of users highlighting specific behaviors under the homophilic assumption (users who act similarly are likely to share similar ideological beliefs).
\textit{Inference Architecture}: a classifier used to infer user connections in a latent space.

\section{Related Work}
\label{ide:sec:prior-work}
Two corpora relate to our study; ideology detection and psychosocial asymmetries.
Our primary concern, for the prior, is pipeline delineation criteria and, for the latter, is evidence bases for hypotheses.

\subsection{Ideology Detection Delineation}
Ideology detection is becoming popular and relevant for researchers and practitioners across the computer, social, and political sciences. 
We delineate prior work by population scope, homophilic lenses, and ideological proxies.

\sh{Population scope} describes \emph{who the technique applies to?} 
Many works limit their scope to a population subset: legislators, elites \cite{xiao2020timme}, the politically active \cite{darwish2020unsupervised}, or everyone \cite{samih2021few}.
Subsets offer clearer ground truth and easier inference, but lack representativeness of the population, leading to biases when applied broadly \cite{alkiek2022classification,cohen2013classifying} and constraining the representativeness of correlational analyses \cite{alizadeh2019psychology}.
This work applies to all users, providing representative downstream analysis.

\sh{Homophilic lenses} describe \emph{which features are utilized to infer ideology?} 
Underlying detection is the homophilic assumption -- people who act similarly are likely to share similar ideological beliefs.
Prior works operationalize this via several lenses: content (including metadata, images \cite{xi2020understanding}, and text \cite{preoctiuc2017beyond}), network (such as followership and resharing \cite{xiao2020timme}), or a combination \cite{chakraborty2022fast}.
In political science, the modus operandi is Ideal Point Estimation \cite{poole1985spatial} using homophily via legislator voting behavior. 
Ideal Point Estimation techniques are largely unsupervised and rely on distinct behavioral patterns but are used in most political science ideology measurement work \cite{gu2016ideology,o2023measurement}. 
In particular, \citet{barbera2015birds} utilize the \emph{following of politicians} on Twitter to estimate user ideal points, and their work is employed in correlation analysis \cite{badaan2023ideological}. 
Given the host of behaviors that portray ideology, novel lenses continue to emerge, including media sharing \cite{cann2021ideological, eady2020news}, and community participation \cite{ravi2022classifying}.
Prior works commonly engineer salient lenses and seek their optimal combination \cite{darwish2020unsupervised, aldayel2019your}; 
however the complexity of data context, inference architecture, and ideological proxy choices often make the conclusions unclear.
\revR{For example, \citet{darwish2020unsupervised} recommend a retweet lens, while \citet{aldayel2019your} recommend a network and lexical lens combination.}
The ideological salience of lenses and their combinations is not our work's focus.
We implement three homophilic lenses previously shown to be ideologically salient, to limit interaction effects with ideological proxies concerns. 

\sh{Ideological Proxy} describes \emph{what is the source of ideological knowledge?}
Prior work utilizes three paradigms for detection: supervised, unsupervised, and weak supervision. 
Each employs distinct ideological knowledge sources-- dubbed \emph{ideological proxies}.
In this study, we focus on both the proxies' performance and their expert annotation labor requirement when used across multiple contexts.
We delineate proxies by 
(1) the extent to which they require expert annotation, 
(2) are transferable to different contexts, and 
(3) how well they represent \emph{true} ideology.
These criteria describe how well proxies generalize to arbitrary datasets and how much manual effort is required for switching contexts.

Direct user annotation for supervised learning \cite{thomas2022reclaim,xiao2020timme} is simple, the most representative, and accommodates fine-grained distinctions between ideologies \cite{liu2023ideology}; however, it is also the most restrictive, requiring laborious expert evaluation of users, across every new context.
Conversely, unsupervised approaches need little annotation and, in theory, are applicable in any context.
Some apply embedding and clustering techniques \cite{darwish2020unsupervised, samih2021few,rashed2021embeddings}. 
Others utilize matrix factorization to jointly learn representations of users and their behaviors \cite{lai2022estimating,lahoti2018joint}. 
These methods are not robust in practice, require highly polarized contexts, fail on homogenous user sets, and depend heavily on lenses. 
Furthermore, they require expert knowledge in post-analysis (e.g., identifying clusters) \cite{darwish2020unsupervised}, and clusters do not always align with ideology.
Weak supervision trades-off between the high labor of supervised and the instability of unsupervised methods.
It employs an ideological proxy, a user behavior strongly correlated with ideology.
Prior work utilizes a range of ideological proxies, including; politically-charged hashtags \cite{rizoiu2018debatenight}, political party relationships \cite{eady2020news}, politician relationships, community participation \cite{lai2022estimating}, and news media sharing \cite{jiang2022retweet,badawy2019falls,Bailo2023}. 
We assess proxies' labor minimization and context transferability qualitatively in \cref{ide:subsec:ideological-proxies} and assess their representativity quantitatively in \cref{ide:subsec:preduction-ground-truth}.

\sh{Related Ideology Detection.}
\citet{jiang2022retweet} use text and retweet features, and a combined media-hashtag proxy which they validate. However, they limit scope to active users who retweet and require hashtag proxy labeling.

\subsection{Psychosocial Profiling of Ideological Groups.}
Many social science works detail the nuanced profiles of fine-grained ideological groups and highlight the asymmetries between ideologies \cite{tomkins1963left,jost2017asymmetries, rao2017red, rao2022partisan}, often requiring painstaking surveys and ethnographic inquiry. 
\revR{
In this work, we supply large-scale online evidence for hypotheses surrounding psychosocial asymmetries of ideologies, relating to morality, grievance, nationalism, and dichotomous thinking.

\sh{Morality.} 
Moral Foundations Theory \cite{graham2009liberals} is an explanation of moral values variations between liberals and conservatives (see \cref{ide:sec:preliminaries}).
Despite its support in psychological survey data \cite{graham2009liberals}, and a handful of online studies \cite{reiter2021studying, mokhberian2020moral}, online social data inconsistently supports this explanation \cite{wang2021moral,alizadeh2019psychology}.

\sh{Grievance and Grudge} are linked to extreme ideologies in psychological theory;\citet{van2021grievance} link grievance to extremism, and \citet{Stankov2021} link grudge to the far-right. 

\sh{Nationalism} is definitionally associated with right-wing politicians.
Prior work has shown that flags are associated with nationalism \cite{kemmelmeier2008sowing}, emojis hold identity and semantics information \cite{li2020emoji}, and that flag emojis are significant in right-leaning political communication \cite{kariryaa2022role}.
However, this research is limited to politicians in a US context.

\sh{Dichotomous Thinking} is a cognitive distortion in people with internalizing disorders, is tied to language \cite{bathina2021individuals}, and is associated with the right \cite{meyer2020political}.
}

Our concern is evaluating hypotheses in large-scale online populations in various contexts.
Accordingly, we limit our scope to automated techniques using online metadata alone.
Prior work, online, analyses left-right \cite{reiter2021studying} or extremist asymmetries, but rarely both \cite{alizadeh2019psychology}. Additionally, they analyze small and non-representative samples. 
This work analyzes left, right, and far-right ideologies in several large-scale online contexts.

\section{Preliminaries}
\label{ide:sec:preliminaries}
Our study relies on several techniques from prior work. 

\sh{Encoding Techniques} are employed to implement homophilic lenses; the Universal Sentence Encoder (USE) \citep{cer2018universal} for our lexical lens (a mature, off-the-shelf, transformer-based model), Term-Frequency Inverse Document Frequency (TF-IDF) for our hashtag lens, and a multi-hot encoding for our resharing lens.
We utilize simple encoding techniques, as they are not our work's main focus.

\sh{Inference Architecture Implementation.}
We use LightGBM \citep{ke2017lightgbm} -- an efficient tree-based classifier -- and FlaML \cite{wang2021flaml}, a system that infers hyperparameters based on dataset characteristics in pipelines.

\sh{Moral Foundations Theory (MFT)} \cite{graham2009liberals} explains variations in moral reasoning through five modular foundations.It espouses that liberals express individualizing foundations (care and fairness) while conservatives express binding foundations (loyalty, authority, and sanctity) relatively more.
We characterize users' language with FrameAxis \cite{kwak2021frameaxis}, a dictionary embedding technique, to identify a user's value for each foundation.
It supplies measures, \emph{bias} and \emph{intensity}.
Importantly, dictionary-embeddings are generally a refinement over dictionaries alone, particularly for smaller documents, however they do not capture the complexities of human language. 
For example, such approaches will not handle negations (for example, ``I do not care'') and do not consider the context around word usage.
Large-language model (LLM) approaches may improve these deficits; however, LLMs introduce their own complexities \cite{liscio2023does} and the dictionary/embeddings approaches are better validated.

\sh{Grievance Dictionary} \citep{van2021grievance} is curated for threat assessment, including categories such as fixation, violence, and paranoia. 
It is validated on social media data, and provides features for distinguishing extremist texts.

\sh{State-of-The-art Baselines.} 
In \cref{ide:sec:results}, we compare our approach to three state-of-the-art detection approaches.
\emph{UUS} \citep{darwish2020unsupervised} encodes the $k$ most active users, applies dimensionality reduction, clusters these embeddings, and assigns clusters stances via expert annotation. 
The authors tune parameters including; $k$, features (based on retweets, retweeted accounts, and hashtags), dimensionality reduction schemes, and clustering schemes.
They recommend encoding $1000$ users via retweets, then applying UMAP and Mean-Shift.
\emph{UUS+} \citep{samih2021few} extends \emph{UUS} by finetuning BERT with \emph{UUS}-labels; applying it to remaining users.
Finally, \emph{TIMME} \citep{xiao2020timme} is a supervised multi-task multi-relation deep graph method using five user relationships to embed and classify users.

\section{Ideology Framework and Implementation}
\label{ide:sec:methodology}
In this section, we describe our ideology pipeline framework in two parts; \cref{ide:subsec:pipeline-cosntruction} partitions pipelines into four components 
and \cref{ide:subsec:ideological-proxies,ide:subsec:feature-lenses} provides component implementation details.

\subsection{Pipeline Constructions Framework}
\label{ide:subsec:pipeline-cosntruction}
In this section, we abstract four components of ideology detection, shown in \cref{ide:fig:schema}: the dataset, ideological proxy, homophilic lens, and inference architecture.

\sh{The dataset} is a set of unlabelled users and their activity metadata within a context.
It has an underappreciated effect on observed pipeline performance. \cref{ide:sec:datasets} discusses classification \emph{difficulty} and introduces our evaluation datasets.

\sh{The ideological proxy} infuses ideological knowledge via weak supervision. 
A user subset is labeled (left, right, or far-right) via ideology-correlated behaviors, such as sharing hashtags, following political parties, endorsing politicians, or sharing news media. See \cref{ide:subsec:ideological-proxies} for details.

\sh{The homophilic lens} characterises ideologically salient user similarity. \cref{ide:subsec:feature-lenses} describes three homophilic lenses: the lexical lens, the hashtag, and the resharing lens.

\sh{The inference architecture} propagates labels from a user subset to the remaining unlabelled users.
We train a classifier on the ideology-proxy-labeled users represented via homophilic lenses.
We use LightGBM with FlaML as our classifier\footnote{The hyperparameter \texttt{n\_estimators} is inferred for the far-right detection due to the sparsity of labeled users;
it is fixed to $200$ for left-right detection to prevent overfitting. 
We set the \texttt{is\_unbalance} flag due to label imbalance.}.
The remainder of this section enumerates the ideological proxies (\cref{ide:subsec:ideological-proxies}) and homophilic lenses (\cref{ide:subsec:feature-lenses}) evaluated, and their implementations.

\subsection{Implementating Ideological Proxies}
\label{ide:subsec:ideological-proxies}

Here, we qualitatively compare proxies and describe the implementations of the proxies evaluated in our study.

\sh{Proxy Qualitative Comparison.}
We conduct an assessment of proxies, based on their utility for practitioners.
Based on our reading of the thematic review presented in \cref{ide:sec:prior-work}, we qualitatively build three criteria to assess each proxy. 
The criteria are designed to partially order proxies as a guide to practitioners.
Therefore, we apply a four-star rating (one star is lower) for each criterion, as shown in \cref{ide:tab:proxy_comparison}.

The first criterion we construct is \emph{labor minimization} (AL) defined as the extent to which expert labor is required to generate the proxy.
Proxies which require human experts to perform the entire contruction will score one star, whereas an approach with no human intervention scores four stars.
The second criterion is \emph{context transferability}\footnote{Note that context transferability has a multiplier effect on annotation labor since a failure to transfer requires reannotation.} (CT), defined as the number and diversity of contexts in which a proxy can be applied.
If a proxy is only available in a given context it will score one star, whereas if the proxy is available with no restrictions, it will score four stars. 
The third criterion is \emph{availability to practitioners}\footnote{We do not discount prior work labor. 
However, we recognize that availability differs, independently of ideology tasks, and proxies' maintenance should be considered in practitioner guidelines.} (AV), defined as the extent to which a proxy or its ingredients are openly available, either for ideology detection or independent tasks.

\begin{table}[]
    \centering
    \label{ide:tab:proxy_comparison}
    \small
    \begin{tabularx}{\columnwidth}{l|XXXX}
      \toprule
               Proxy & AL  & CT & AV  \\
      \midrule
        \hashtags    &  *  & *   & * \\
        \communityparticipation    &  **  & **   & * \\
        \politicianHOP    &  **  & **   & ** \\
        \politicalParties    &  ***  & **   & *** \\
        \media    &  ***  & ****   & **** \\
      \bottomrule
    \end{tabularx}
    \caption{
      \textbf{Ideology Proxy Qualitative Comparison} for application by practitioners based on three-part criteria; annotation labor minimization (AL), context transferability (CT), and Availability (AV). Criteria are rated out of four-stars.
}
  \end{table}
\hashtags shared is commonly used as a proxy, but requires domain knowledge and is time-consuming to generate (one star on AL), and generally requires reannotation for every dataset (* for CT). 
Furthermore, not all social media platforms use hashtags therefore it has a low availability (* for AV).
\communityparticipation uses user activity in ideological communities (e.g., subreddit posting). 
The communities tend to be fewer and more persistent (** on AL) and there is some detectable overlap of the communities between platforms (** on CT).
However, they are unavailable on some platforms (e.g., Twitter/X) and are inconsistent across countries (* on AV). 
Furthermore, it requires experts for annotation, and datasets linking communities to ideologies are few.
\politicalParties and \politicianHOP leverage databases of political parties and politicians with their online profiles, which are intermittently available (** on AV).
Such databases are usually country- and period-specific -- political parties emerge, change, and become relegated in time.
The advantage of these proxies is their stability and non-ambivalent nature during the studied context (** on CT).
Furthermore, databases do not encode all ideologically relevant information, such as the lean of parties or specific politicians, requiring an expert instead (** and *** on AL, respectively).
\media proxies utilize users sharing news media, which often have known political slants. 
They leverage available and well-maintained data on media slants (*** on AL), which have intrinsic value in communication studies, the news ideology detection task, and general consumer value. 
There is strong evidence linking news readership \cite{garimella2021political, bakshy2015exposure} and sharing \cite{an2014partisan} to ideology.
Media slants are fairly consistent across time, media-sharing behaviors occur on most platforms (**** on AV), and media tend to be ideologically consistent across topics (**** on CT).
There are limitations to the media proxy (see \cref{ide:sec:Conclusion}), but it outperforms its alternatives in terms of annotation labor, context transferability, and general availability.

\paragraph{Left-Right ideological proxies.} 
We build four proxies.

\sh{\hashtags} proxy requires experts to code hashtags.
We qualitatively inspect the $1,000$ most common hashtags in our datasets and label their political lean; 
$-1$ if left-leaning, $0$ if non-partisan, and $1$ if right-leaning. 
We quantify a user's political lean as the mean of the labeled hashtags they emit and their ideology label as the sign of this lean.

\sh{\politicalParties} proxy requires collecting the followers of the major political parties' online accounts for each target country (i.e., Australia and USA). 
We code the political parties by their ideology.
The users in the dataset who follow a single party receive the party's ideology label. 

\sh{\politicianHOP} proxy requires a dataset of politicians, their political affiliations, and social media handles.
We use the Twitter Parliamentarian Database \cite{van2020twitter}. 
We code the politicians using their party's ideology (where independents are excluded).
Note that independents' exclusion reduces the proxy representativity, but this is preferable to manually labeling all independents.
We label users who retweet politicians using the majority vote of the politicians' ideologies.

\sh{\urlLR} (Media Publication Proxy) requires a dataset of media websites with their political slants. 
We utilize an extensive survey \cite{park2021digital, newman2021reuters} of news consumption behavior within English-speaking countries (Australia, New Zealand, UK and the USA), collected in 2020 and 2021 by Reuters.
Participants indicated the news media they read and self-reported their political leaning ranging from $-3$ (extreme left) to $3$ (extreme right).
We compute a publication's slant
as the weighted mean political lean of the participants who consume that publication, where each participant is weighted by the inverse number of publications they consume. 
Since countries' perspectives on what constitutes left- and right-leaning differ, we calibrate scores across countries with the AllSides Media Bias Ratings \cite{allsides}.
We encode the ratings' five-point scale onto a numerical scale from $-1$ to $1$.
We align each country's scores, minimizing the sum of squared differences between a country's scores and AllSides scores for overlapping publications.
Finally, we generate slant scores for each publication as the average slant over all countries and years. 
We associate publications (and their slants) with their website domains, averaging where a domain is shared.
We present the media organizations and their constructed slant scores in online appendix \citep{appendix}.
We compute a user's political lean as the average lean of the media domains they share and their ideology label as the sign of this lean.

\paragraph{Far-Right ideology proxies.}
We build two proxies.

\sh{\urlFR} is constructed from the media slant scores of mainstream media built for \urlLR.
Next, we label users `far-right' if their political lean exceeds $0.5$ or as `moderate' otherwise.

\sh{\mbfc} is constructed from the Media Bias Fact Check \cite{mbfc} dataset, including both media slant and veracity, and containing conspiratorial and fake news sources.
We label users sharing the right-most media category as `far-right'.

\subsection{Homophilic Lenses}
\label{ide:subsec:feature-lenses}
Homophily is the tendency of similar users to be similar \cite{mcpherson2001birds} and is commonly assumed in ideology detection. 
A \emph{homophilic lens} is a user embedding that encodes ideologically relevant information.
Here we convert content about user behavior into numerical vectors.
This section details three lenses.

\sh{Lexical lens (USE).}
Language is a strong indicator of one's political ideology \cite{cichocka2016grammar}; since a sociolect is formed through associations with others. 

\sh{Hashtag lens (HT).} 
Hashtags signal users' interests and the discussion topics they participate in \cite{bode2013mapping}. 

\sh{Resharing lens (RT).} 
Resharing is a signal of endorsement \cite{metaxas2015retweets}. 
We assume users endorsing the same people likely share similar ideologies \cite{van2021political}.

\sh{Implementation.}
For the \emph{lexical lens}, we preprocess text, to prevent potential data leaks, by removing URLs, hashtags, and mentions. 
We concatenate each user's tweets and encode them as $512$ dimensional vectors via the universal sentence encoder (USE) \cite{cer2018universal}. 
The encoder choice is arbitrary and based on its prior user in literature for social media-originating text \cite{rashed2021embeddings}. For the \emph{hashtag lens}, we use the Term-Frequency Inverse Document Frequency (TF-IDF) of users (i.e., documents) via hashtags (i.e., words) they use if used at least $10$ times.
TF-IDF is a refinement over the bag-of-words model that weights terms used by their occurrence within a corpus, providing a simple but salient vector representation.
Finally, for the \emph{resharing lens}, we generate a multi-hot encoding for users based on the $1000$ most reshared posts. 
We represent a user $u_i$ as $h_i \in \mathbb{R}^{1000}$, where $h_i[j] = 1$ if $u_i$ reshares the $j$th most reshared post ($h_i[j] = 0$ otherwise). 
In summary, there are three representations of users; lexical $\mathbb{R}^{512}$, hashtag $\mathbb{R}^{|\text{\#hashtags}|}$, reshare $\mathbb{R}^{1000}$.

\section{Contexts, Datasets, and Ideology Labels}
\label{ide:sec:datasets}

This section introduces datasets and their contexts.
\cref{ide:subsec:datasets-context} describes the five datasets, and \cref{ide:subsec:gold-standard} shows how we qualitatively construct ideology ground truth, used to evaluate proxies and pipelines' performance.

\subsection{Contexts and Datasets}
\label{ide:subsec:datasets-context}
\cref{ide:tab:datasets} summarizes the datasets; there are three Australian and two American datasets; one originates from Parler, another is a mixture of Facebook and Twitter, and the remainder are Twitter-based.
In prior work datasets, ideology correlates with explicit user behavior (e.g., discussion topics); this simplifies detection but rarely holds in practice.
Here, we use data where detection is difficult, as one would likely encounter in the wild.
We quantify the \emph{detection difficulty} using Hopkin's statistics~\citep{hopkins1954new} of the lexical lens, indicating the clustering tendency of data, ranging from 1 (highly clustered, easy detection) to 0 (uniformly distributed, difficult detection).
Hopkin's statistic is common measure of clustering tendency, effectively characterizing the probability that embeddings are drawn from a uniform distribution.
We assume that embeddings with a high clustering tendency are easier to classify.
Note clusters do not necessarily align with classes, however they often do in real-world data; baselines, like UUS and UUS+, directly employ this axiom to infer labels (relying heavily on the underlying clustering tendency of the data).
Quantifying detection difficulty of datasets is uncommon in literature and prior work often vary dataset difficulty be construction \cite{macia2008genetic} or require class labels to infer it \cite{lorena2019complex}.
We employ Hopkin's Statistic as a simple quantification of difficulty (which is not the focus of our work).
It is likely related to the decision boundary aspect of classification complexity \cite{lorena2019complex}.
\cref{ide:tab:datasets} shows values $\in [0.14,0.3]$ indicating no clustering tendency.

\begin{table}[tb]
    \label{ide:tab:datasets}

    \centering
    \small
    \setlength{\tabcolsep}{4pt}
    \begin{tabularx}{\columnwidth}{{lrrlr}}
        \toprule
        \textbf{Dataset}                                          & \textbf{\#Users} & \textbf{\#Posts} & \textbf{Country} & \textbf{Hopkins} \\ \midrule
        \qanda                                                    & 103,074          & 768,808          & AUS              & 0.2624           \\
        \ausvotes                                                 & 273,874          & 5,033,982        & AUS              & 0.2445           \\
        \socialsense & 49,442           & 358,292          & AUS              & 0.2591           \\
        \riot{}                                       & 574,281          & 1,067,794        & US               & 0.1490           \\
        \parler{}                                   & 120,048          & 603,820          & US               & 0.3016           \\ \bottomrule
    \end{tabularx}
    \caption{
        \textbf{The datasets used in this work:} 
        source, profiling, and country of origin (AUS and US refer to Australia and USA, respectively). The last column represents the Hopkins statistics \cite{hopkins1954new} for the lexical lens.
    }
\end{table} 
Briefly, the datasets are:
\emph{\qanda} [Twitter/X] surrounding a political panel show with audience questions; \emph{\ausvotes} [Twitter/X] surrounding the 2022 Australian Federal Election; \emph{\socialsense} [Twitter/X and Facebook] \citep{calderon2022opinion} surrounding the Australian Black Summer Bushfires; \emph{\riot} [Twitter] \cite{riot} and \emph{\parler{}} [Parler] \cite{parler} both surrounding the US capitol insurrection. See \cite{appendix} for details.

\subsection{Build a Ground Truth} 
\label{ide:subsec:gold-standard}
We qualitatively annotate a subset of \qanda users to generate both a left-right and far-right ground truth. 

\sh{Left-right ground truth.}
Due to the imbalance and sparsity of some ideological classes\footnote{Predicted label counts show this imbalance \cite{appendix}.}, we employ the proposed pipeline to construct a candidate set of users for manual annotation.
Platforms such as X/Twitter have been shown to lean-left, and the imbalance in datasets (such as Q+A which attracts a left-leaning audience) can be substantial. 
While it can be argued that using the pipeline to generate a ground truth to train future pipelines may skew the data selection, it has advantages over the alternatives. For example, (1) conducting a manual search through a random candidate set and generating a proportionately low-volume of right-leaning users is prohibitively expensive, and 
(2) employing a proxy directly as our ground truth (following the baselines we compare against) defeats the purpose of evaluating the proxies and introduces significant biases.

We generate the candidate set using the following four components;
(1) we select each of the four proxies (\hashtags, \politicalParties, \politicianHOP, \urlLR),
(2) using labels derived from the selected proxy, we train the classifier to predict user labels (since even proxy do not necessarily produce sufficient volumes of right-leaning users),
(3) we apply to proxy-trained classifier to the entire \qanda dataset (including those already labelled),
(4) finally, we extract the $100$ left- and $100$ right-leaning users with the highest classifier confidence (estimated through the classifier sigmoid scores).
We collect the pool of $800$ users in one set, deduplicate, shuffle it, and remove users who are unavailable (either private or suspended).
This results in $695$ users; we sample $200$ users, inspect their profiles and categorize them as left-leaning, right-leaning, far-right, or indeterminable.

Next, two experts manually labeled each profile. 
The experts both had extensive knowledge of the Australian political context, and were native English speakers.
They were given examples of left, right, far-right, and indeterminable user profiles for context.
They were instructed to use any signals of ideological-alignment they observed to make their assessments (see \cite{appendix} for details).
Finally, they were given links to each user profile and instructed to categorize them.
They achieved moderate inter-annotator agreement i.e., Cohen's $\kappa$ of $0.515$
As a result, our left-right ground truth contains $103$ left- and $74$ right-leaning users.

\sh{Far-right ground truth.}
\citet{Bailo2023} snowball sample Australian far-right users, starting with a `seed' user and recovering `lists' (a Twitter feature documenting similar users) they belong to. They intersected the sample with their dataset, manually validated their far-right status, crawled this validated set's followers, and manually coded these too. They obtained $1,496$ users, of which $686$ are in \qanda, and serve as our far-right ground truth.

 \section{Proxy Bias, Baselines, and Validation}
\label{ide:sec:results}

In this section, we first quantify proxy bias (i.e., representativity) and homophilic lens interaction effects, by enumerating all pipeline constructions, in \cref{ide:subsec:optimal_proxy}.
Next, we present a pipeline construction that outperforms three state-of-the-art methods in \cref{ide:subsec:preduction-ground-truth}.
Finally, we evaluate transfer learning across contexts, illustrating `in-context' training superiority, 
and test cross-proxy performance in \cref{ide:subsec:generalization}.

To avoid confusion, \cref{ide:subsec:optimal_proxy} employs both ground-truth and \cref{ide:subsec:preduction-ground-truth} uses the left-right ground-truth constructed in \cref{ide:subsec:gold-standard} for the \qanda dataset.
\cref{ide:subsec:generalization} does not utilize the constructed ground truth. 
In its first segment it trains on labels derived from one proxy and tests on labels derived from another, with fixed dataset \qanda.
In its second segment it trains on users from one dataset and tests on users from another, with fixed proxy \urlLR.

\begin{table}[tbp]
  \centering
  \label{ide:tab:gold_ablation_table}
\setlength\tabcolsep{3pt}
\small
  \begin{tabularx}{\columnwidth}{l|rrrr|rr}
    \toprule
        & \multicolumn{4}{c|}{Left-Right}      & \multicolumn{2}{c}{Far-right} \\ \midrule
        & \multicolumn{1}{l}{\thead{Hashtags}} & \multicolumn{1}{l}{\thead{Party\\ Follow.}} & \multicolumn{1}{l}{\thead{Pol.\\ Endors.}} & \multicolumn{1}{l|}{\thead{L.R.\\ MPP}} & \multicolumn{1}{l}{\thead{F.R.\\ MPP}} & \multicolumn{1}{l}{\thead{MBFC\\ MPP}} \\ 
    USE       & 0.881 & 0.868 & 0.788 & 0.946  & 0.691  & 0.773  \\
    HT        & 0.873 & 0.876 & 0.812 & 0.849  & 0.559  & 0.633  \\ 
    RT        & 0.840  & 0.844 & 0.752 & 0.879  & 0.538  & 0.668  \\ 
    USE+HT    & 0.949 & 0.879 & {\underline{{0.870}}} & 0.939  & {\underline{{0.715}}} & {\underline{{0.785}}} \\ 
    USE+RT    & 0.880  & 0.821 & 0.785 & {\underline{{0.953}}} & 0.666  & 0.762  \\ 
    HT+RT     & 0.904 & {\underline{{0.914}}} & 0.799 & 0.937 & 0.570   & 0.632  \\ 
    \textit{all} & {\underline{{0.950}}} & 0.875 & 0.854 & 0.929 & 0.713      & 0.785    \\ \midrule
Prec. & 0.889 & 0.873 & 0.797 & \textbf{0.892}  & 0.516  & \textbf{0.530}  \\
    Recall    & 0.857 & 0.820 & 0.794 & \textbf{0.902} & 0.540  & \textbf{0.557}  \\
    F1        & 0.855    & 0.821 & 0.766    & \textbf{0.893}     & 0.636     & \textbf{0.720}     \\
    \bottomrule
  \end{tabularx}
  \caption{
    \textbf{Determine the optimal proxy and lens combination.} 
    \textit{(top)} 
    AUC ROC for each combination of lenses (rows) and proxy (columns).
    The underlines show the best lens for a given proxy.
    \textit{(bottom)}
    The precision, recall and macro-F1 for each proxy averaged over all lens combinations.
    The bold show the best-performing proxy.
}
\end{table}
 
\subsection{Quantifying Proxy Bias}
\label{ide:subsec:optimal_proxy}
Here we jointly assess ideological proxy and homophilic lens combinations and their performance against our ground truths, to infer proxy representativity.
The top section of \cref{ide:tab:gold_ablation_table} shows all combinations.
The columns represent proxies, and the rows show the seven possible concatenations of our lens implementations.
We use the respective ground truth for validation and testing in a $50\%:50\%$ split, employing the validation set for threshold calibration (for converting continuous scores to discrete predictions), and removing neutral ideologies from training, as they do not appear in testing. 
Cells show AUC ROC scores for pipelines trained with respective proxy and lens combinations.
A higher AUC ROC score is better with a maximum score of $1$ and a random baseline of $0.5$.
The bottom section of \cref{ide:tab:gold_ablation_table} shows the precision, recall and F1, averaged over all lens combinations. 
The purpose is to quantify how well proxies represent `true' ideology, approximated via our ground truth.

\sh{Results.}
There are two main conclusions.
First, \cref{ide:tab:gold_ablation_table} (bottom) shows that MPP consistently outperforms other proxies for left-right detection.
In order of representativity, we have \urlLR, \hashtags, \politicalParties, and \politicianHOP.
The \mbfc is the most performant for far-right ideology.
\revR{This is significant, as we have shown that media-based proxies are both qualitatively advantageous and optimal for representativity; providing clear guidelines for practitioners.}
Second, \cref{ide:tab:gold_ablation_table} (top) shows that no homophilic lens dominates all others and the best-performing lens combination changes for each proxy.
\revR{This may explain unclear conclusions within the literature, where lens optimization is performed in isolation of other pipeline components (e.g., proxies).}
Despite the lack of a dominating lens, we observe that pipelines containing the lexical lens generally outperform their peers, and USE by itself (first row) has competitive performances.
In addition, USE is the only platform-independent lens.

\subsection{Prediction Performance Against Baselines}
\label{ide:subsec:preduction-ground-truth}

\begin{figure}[tb]
    \centering
    \newcommand\myheight{0.245}
    \includegraphics[height=\myheight\textheight,valign=c]{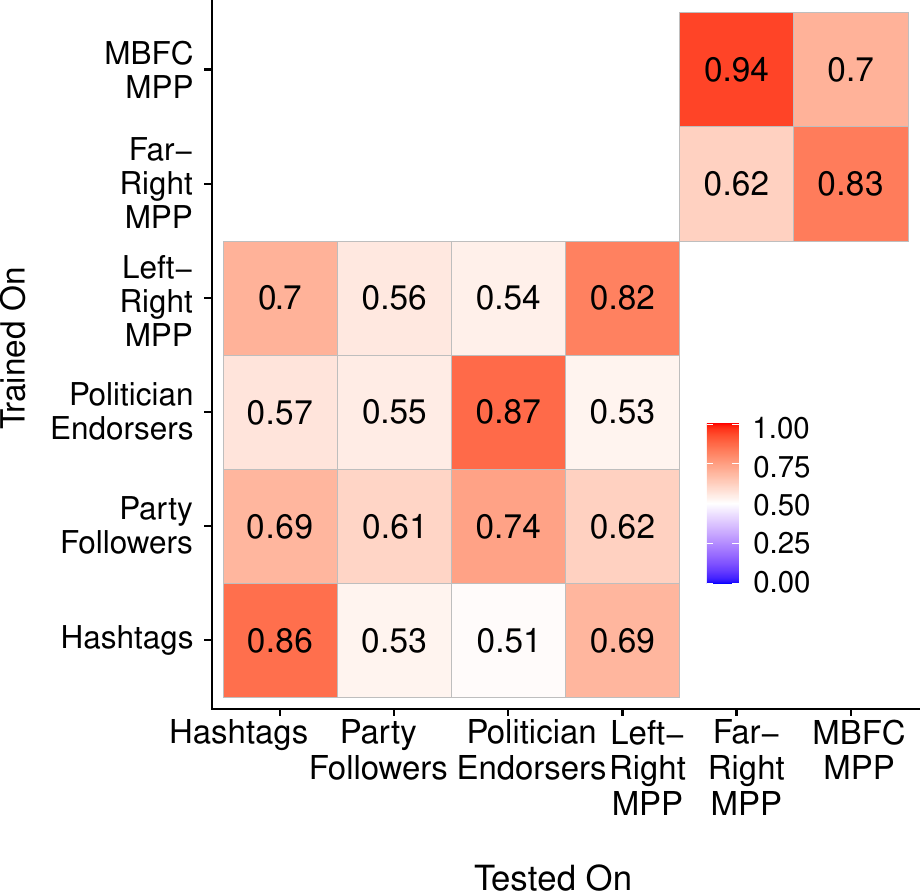}
    \label{ide:subfig:cross_ground_truth}
    \caption{
        \textbf{Self- and cross-proxy generalization.} 
        The AUC ROC of ideology detection on \qanda when trained on one proxy (y-axis) and tested on another (x-axis) for left-right far-right proxies.
    }
\end{figure}

\begin{figure}[tbp]
    \centering
    \newcommand\myheight{0.245}
    \includegraphics[height=\myheight\textheight,valign=c]{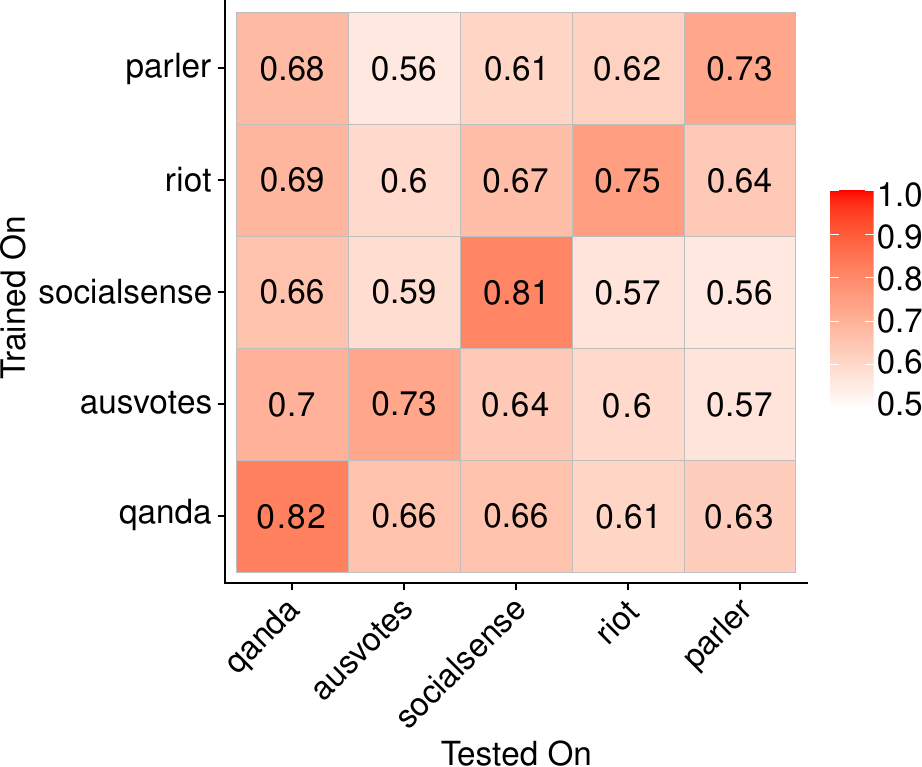}\vphantom{\includegraphics[height=\myheight\textheight,valign=c]{images/cross_gt_performance.pdf}}\label{ide:subfig:cross_dataset}
    \caption{
        \textbf{Context generalization.} 
        AUC ROC of \urlLR trained on one dataset (y-axis) and tested on another (x-axis). 
    }
\end{figure}

\begin{table}[tb]
    \centering
    \label{ide:tab:baselines}
\setlength\tabcolsep{6pt}
    \begin{tabularx}{\columnwidth}{lXXXX}
        \toprule
        Method & \emph{UUS} & \emph{UUS+} & \emph{TIMME} & Ours \\ \midrule
        Macro-F1 & $0.60_{\pm 0.23}$ & $0.61_{\pm~0.26}$ & 0.88 & 0.92 \\
        AUC ROC & -- & $0.76_{\pm~0.15}$ & 0.89 & 0.95 \\
        \bottomrule
    \end{tabularx}    
    \caption{
        \textbf{Baselines.} 
        Left-right classification performance of baselines vs. our pipeline on the ground truth. 
        We report the mean and standard deviation over all setup combinations for \emph{UUS} and \emph{UUS+}. 
        Note that \emph{UUS} does not produce a score, only labels; therefore, AUC ROC cannot be computed for it.
    }
\end{table}
 \sh{Baselines.}
We evaluate a pipeline construction against three state-of-the-art stance detection techniques: \emph{UUS} \cite{darwish2020unsupervised}, \emph{UUS+} \cite{samih2021few}, and \emph{TIMME} \cite{xiao2020timme} -- detailed in \cref{ide:sec:prior-work}.
For \emph{UUS}, the authors' recommended setup (UMAP+Mean-Shift, retweet features, and $1000$ active users) does not produce any clusters on \qanda.
To render UUS competitive, we enumerate the setups similarly to their work. 
We fix the dimensionality reduction to \texttt{UMAP} and clustering to \texttt{Mean-Shift} following their recommended setup.
We use the default \textit{scikit-learn} settings (\textit{n\_neighbors=15}, \textit{min\_dist=0.1}, \textit{n\_components=2}, \textit{metric=cosine}), and do not enumerate different hyperparameters to (1) faithfully replicate their work, and (2) simulate the experience of a time-poor practitioner.
Furthermore, we implement setups for every combination of features (retweets, retweeted accounts, and hashtags) and number of active users ($500$, $1000$, and $5000$).
In addition, \emph{UUS} only reports the most active users' labels; however, our ground truth users are not the most active. 
Instead, we use UMAP and Mean-Shift inference methods to acquire labels for these users.
For \emph{UUS+}, we use the same set of \emph{UUS} setups.
Following the authors, we utilize BERT\textsubscript{base multilingual}, using the HuggingFace implementations with PyTorch.
We fine-tune BERT by adding a fully-connected dense layer followed by a softmax output layer.  
We minimize the cross-entropy loss over the training data.
As it is not specified by the authors, we choose to fine-tune for $10$ epochs (a sufficient quantity for our data volume).
Finally, for \emph{TIMME}, we use all relations except the followership network, which is prohibitive to acquire.

\sh{Predicting human-annotated ideology.}
We evaluate performance using the left-right ground truth (see \cref{ide:subsec:gold-standard}) with a 5-fold cross-validation (where applicable).
For this task, we use the pipeline constructed from the \urlLR and the USE+RT homophilic lens (the best-performing combination from \cref{ide:tab:gold_ablation_table}).
\cref{ide:tab:baselines} shows the F1-macro and AUC ROC scores for each technique.
We make several observations.
First, our approach consistently outperforms all baselines, with the next best being \emph{TIMME}.
Second, \emph{UUS} and \emph{UUS+} show low mean performance and high standard deviation.
Most setups failed to cluster users and were removed before computing the mean and standard deviation. Furthermore, the clusters required an expert for labeling. 
\revR{Our pipeline construction has practical advantages over these baselines and outperforms them.}

\subsection{Cross Proxy and Context Generalization}
\label{ide:subsec:generalization}
\sh{Cross Proxy Generalization.}
Here, we characterize the robustness of ideological proxies through their self- and cross-consistency.
\emph{Self-consistency} indicates how well the pipeline predictions trained with a given proxy align with the same proxy on a test set.
We evaluate self-consistency using a 5-fold cross-validation.
\emph{Cross-consistency} indicates that two proxies capture similar ideological signals.
We evaluate the directed cross-consistency of a source $\longrightarrow$ target proxy by deploying a pipeline with the source proxy to predict the ideology of every user in the \qanda dataset and testing against the ideology labels set by the target proxy.
We report the performance over users whom the target proxy labels, and use a one-vs-one scheme to adjust to the multiclass setting.
For a given proxy, we deploy the pipeline with the best lens combination as per \cref{ide:tab:gold_ablation_table}.

\cref{ide:subfig:cross_ground_truth} shows the AUC ROC performance for every pair of source $\longrightarrow$ target proxy for both left-right and far-right ideology detection.
\emph{The self-consistency} (main diagonal) is high for all left-right pipelines, except \politicalParties.
It is worth noting, \politicianHOP has high self-consistency but a low prediction performance against the ground truth (see \cref{ide:tab:gold_ablation_table}).
\revR{This suggests that politician endorsement behavior is distinct from prototypical ideological behavior.}
Note, far-right proxies have relatively low self-consistency, perhaps due to the sparsity of far-right users.

\cref{ide:subfig:cross_ground_truth} shows \emph{cross-consistency} of left-right pipelines is relatively low, except for \urlLR and \hashtags. 
This supports prior work \cite{cohen2013classifying, alkiek2022classification} arguing that different proxies confer diverse ideology prototypes. 
The \urlLR and \hashtags proxies generalize well to each other and the ground truth (see \cref{ide:tab:gold_ablation_table}), suggesting they accurately represent \emph{true ideology}.
Both far-right proxies generalize well on each other, but their performance on the ground truth is relatively weak.
This indicates they represent similar behaviors not fully aligned with ideology.

\sh{In-context Dominance.}
Researchers often implicitly suggest political signals from one context transfer to others. 
Here we demonstrate the importance of `in-context' training.
Each dataset is typically associated with a distinct context (see \cref{ide:subsec:pipeline-cosntruction}).
We evaluate transfer-learning across contexts by training a pipeline (constructed with the \urlLR proxy and the USE+RT lenses) on one dataset and testing on another dataset.
\cref{ide:subfig:cross_dataset} shows the 5-fold cross-validation AUC ROC performance of left-right ideology detection for every pair of datasets.
Intuitively, models perform best when trained and tested on the same dataset (i.e., in-context).
However, we observe a significant performance drop-off with transfer learning (off-diagonal).
Despite this, we see relatively better transfer learning between contexts that share traits.
Models trained in Australian contexts perform better when tested within the Australian context, and noticeably underperform when tested in US contexts. 
Moreover, a further reduction is observed when training or testing with the Parler dataset (i.e., a different social platform context).
These observations indicate that signals of ideology differ between contexts. 
While transfer learning performs better in similar contexts, `in-context' training is significantly more effective.

\section{Psychosocial Analysis of Ideology Cohorts}
\label{ide:sec:psychosocial-results}

\newcommand{\rohhead}[1]{\rotatebox{90}{\centering \raggedleft{\thead{#1}}}}

\begin{figure*}[tbp]
    \centering
    \newcommand\myheight{0.2339}
    \subfloat[]{
        \includegraphics[height=\myheight\textheight]{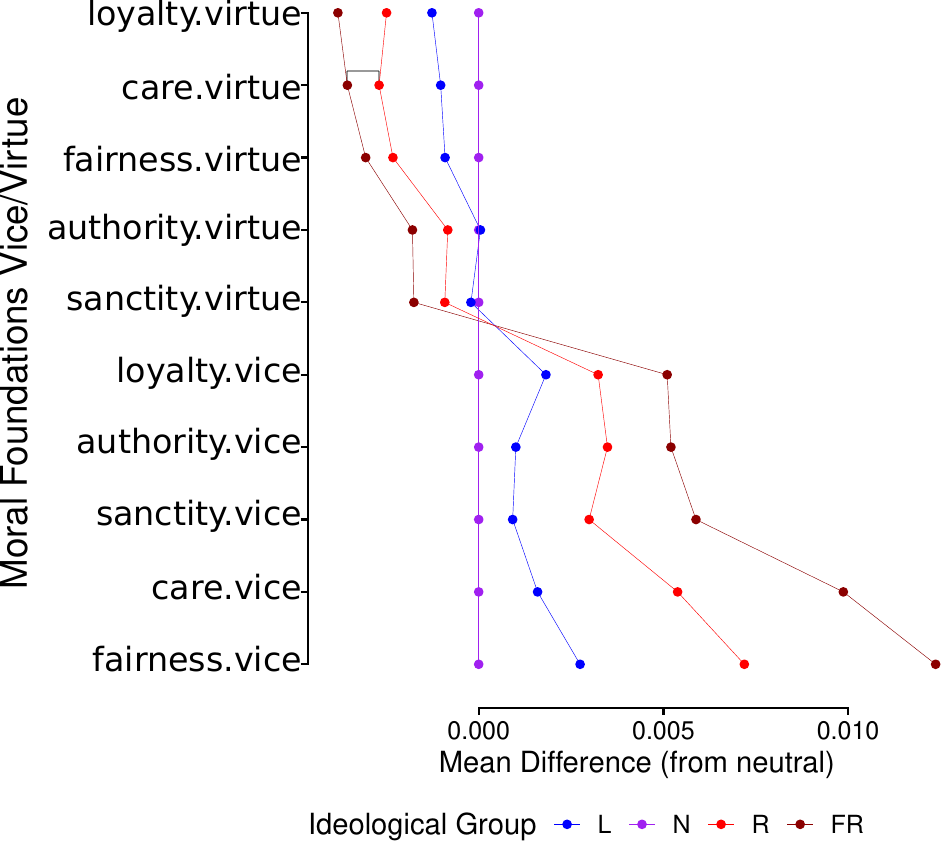}\label{ide:subfig:vice-virtue-socialsense}
    }\subfloat[]{
        \includegraphics[height=\myheight\textheight]{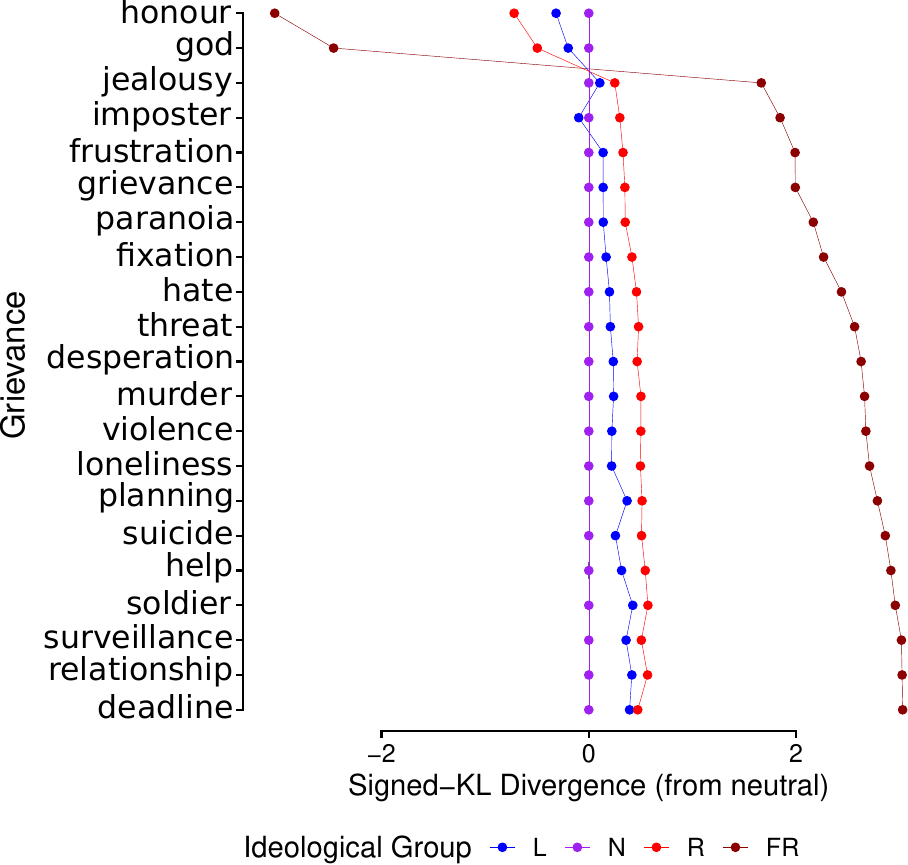}\label{ide:subfig:grievances-ausvotes}
    }\subfloat[]{
\includegraphics[height=\myheight\textheight]{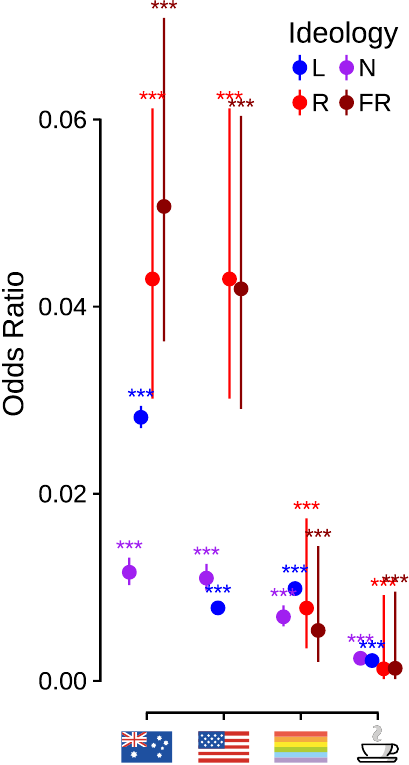}\label{ide:subfig:emoji-nationalism}
    }\subfloat[]{
\includegraphics[height=\myheight\textheight]{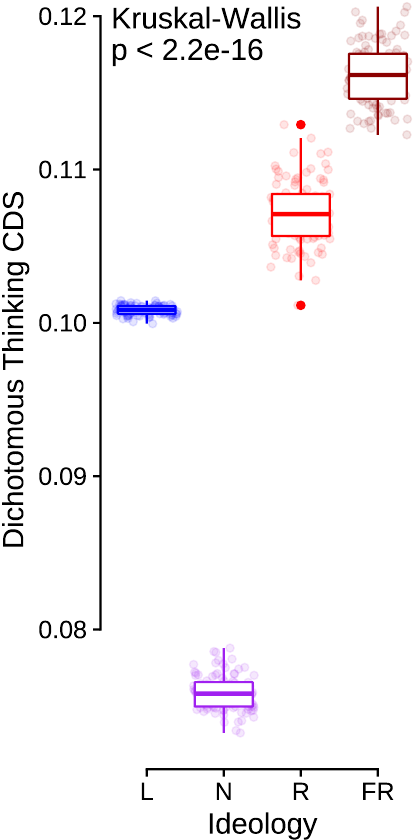}\label{ide:subfig:dichotomous-thinking}
    }\caption{
        \textbf{(a)(b) Distribution of psychosocial properties} for ideological groups for \qanda and \ausvotes, respectively.
        Line color represents ideological groups, and the y-axis shows psychosocial categories.
\emph{(a) Vices-Virtues.} The x-axis is the mean difference for each ideological group from neutral, for Moral Foundations vice and virtue categories.
        \emph{(b) Grievances.} The x-axis is the signed-KL divergence of each group ideological group from neutral for grievance categories.
        (c) \textbf{Emoji Nationalism.} The odds (y-axis) of observing an emoji (x-axis) for a user given their ideological group (color), for \qanda. The odds are determined via logistic regression with no reference group.
(d) \textbf{Dichotomous Thinking.} The bootstrapped prevalence distribution of dichotomous thinking CDS (y-axis) in tweets by users from ideological groups (x-axis), for \qanda.
    }
    \label{ide:fig:set-two}
\end{figure*}
 In this section, we test four hypothesis sets for psychosocial asymmetries of ideologies, relating to morality, grievance, nationalism, and dichotomous thinking. 
This serves two purposes: an application case study for practitioners and to supply online evidence bases for conclusions of prior work. 
We use pipelines constructed from the \mbfc and \urlLR proxies, alongside the USE lens for its applicability across all datasets. The first pipeline labels users as `far-right'. If users are not labeled `far-right', the second pipeline assigns them as `left', `neutral', or `right'. 
For most analysis below, we highlight results on a single dataset, however we produce the relevant plots for all datasets and label distributions in the supplementary material \cite{appendix}.

\sh{Testing Moral Foundations.}
We begin by evaluating MFT hypotheses.
There are five hypotheses relating to individualizing (liberal) and binding (conservative) foundations.
We use a Wilcox Rank Sign Test (95\%), with Holm adjustment for family-wise error, to evaluate the support for each moral foundations hypothesis in each dataset.
We test these hypotheses with the \emph{bias} and \emph{intensity} measures and both ``left vs. right'' and ``left vs. far-right'' (i.e., each combination has four hypotheses).
\cref{ide:tab:hypotheses_table} shows the number of statistically significant tests for each moral foundations hypothesis in each dataset. 
Overall, only $46\%$ of hypotheses are supported, marginally favoring the individualizing over the binding hypotheses.  
This \revR{inconsistency}, seen in prior work \cite{wang2021moral,thomas2022reclaim}, suggests MFT applies differently online than it does offline.
\begin{table}[tb]
\centering
\label{ide:tab:hypotheses_table}
\begin{tabularx}{\columnwidth}{l|XXXXX|r}
    \toprule
    & \rohhead{\qanda} & \rohhead{\ausvotes} & \rohhead{\socialsense} & \rohhead{Riot} & \rohhead{Parler} & Total\\
    \midrule
    Fairness & 2 & 2 & 2 & 2 & 2 & $\sfrac{10}{\gr{20}}$\\
    Care & 2 & 4 & 3 & 1 & 3 & $\sfrac{13}{\gr{20}}$\\
    Loyalty & 2 & 0 & 1 & 1 & 2 & $\sfrac{6}{\gr{20}}$\\
    Authority & 2 & 1 & 2 & 2 & 2 & $\sfrac{9}{\gr{20}}$\\
    Sanctity & 2 & 0 & 1 & 2 & 3 & $\sfrac{8}{\gr{20}}$\\
    \addlinespace \midrule
    Total & $\sfrac{10}{\gr{20}}$ & $\sfrac{7}{\gr{20}}$ & $\sfrac{9}{\gr{20}}$ & $\sfrac{8}{\gr{20}}$ & $\sfrac{12}{\gr{20}}$ & $\sfrac{46}{\gr{100}}$\\
    \bottomrule
\end{tabularx}
\caption{
    \textbf{Moral Foundations Hypotheses testing.} 
    The number of times the MFT hypotheses tests are significant for each foundation (rows) and dataset (columns).
}
\end{table}

Next, given the lack of support for MFT, we test an alternative hypothesis, that \emph{right-leaning users, relative to left-leaning users, exhibit vice over virtue foundations}.
For each moral foundation, we assign each user a virtue/vice score equal to their intensity, if their bias is positive/negative, respectively. 
This segregates the population into vice or virtue users. 
In \cref{ide:subfig:vice-virtue-socialsense}, we plot each foundation's mean vice and virtue scores for each ideological group in the \qanda dataset. 
We observe that a significant proportion of right-leaning users partake in the language of vice rather than virtue compared to left-leaning users. 
We apply the Wilcox Rank Sign Test (95\%) between the means of ideological groups, for each category, and find all are significantly different\footnote{Except between the right and far-right in the \emph{care-virtue} category, which is irrelevant to our conclusions}.
We show that this is relatively consistent across all datasets in the online appendix \cite{appendix}, 
This provides a consistent moral asymmetry in the online context.

\sh{Testing Extremists' association with Grievance.}
Early signals of extremism are of particular concern to national security and law enforcement practitioners. 
Prior work suggests that \emph{extreme ideologies hold more grievance and grudge beliefs than moderates.}
We use the Grievance dictionary \cite{van2021grievance} to quantify users' grievance and grudge language. 
In \cref{ide:subfig:grievances-ausvotes}, we plot the Kuller-Leibach divergence (signed by mean difference) between the distribution of each ideological group from the neutral group for each category with the \ausvotes dataset.
We apply the Kruskal-Wallis Test (95\%) between ideological groups, for each category, and find all are significantly different.
We observe that the far-right users differ significantly from the other ideologies in all categories and generally use more grievance language. 
Notably, in the \ausvotes dataset, the far-right users use \emph{honor} and \emph{god} type language less than other groups. 
In the online appendix \cite{appendix}, we show that this hypothesis holds for most datasets.
A takeaway for practitioners is that far-right language and threat assessment indicators overlap, suggesting a method to build effective public safety tools.

\sh{Testing Nationalism via Emoji.} 
Here we add online evidence that \emph{the right-wing are associated with nationalism} via emojis.
This hypothesis is widely accepted (and definitional), and supporting it validates our inferred ideologies.

\cref{ide:subfig:emoji-nationalism} shows the odds of observing an emoji, given a user's ideological group in \qanda. 
Point ranges indicate the 95\% confidence interval.
The significance of the emoji in predicting ideological groups, via the Wald Test, is indicated with stars.
We make several observations.
First, \emojiAUS is used more by ideological groups than neutral users.
Second, the right (and far-right) use \emojiAUS and \emojiUS significantly more than the left.
Third, \emojiRAINBOW is used marginally more by the left than other groups.
Finally, we include \emojiCOFFEE as a control (showing no associations with any ideology). 
We conclude that nationalism, via national flags, is associated with our inferred right-leaning ideologies. The use of \emojiUS could be evidence of imported ideology from America to Australia. \emojiRAINBOW is only marginally associated with the left.

\sh{Testing Dichotomous Thinking.}
Recent work suggests \emph{the right-ideologies applying black-and-white thinking relatively more than left-wing ideologies}.
Following \cite{bathina2021individuals}, we match n-grams relating to cognitive distortions schema (CDS) in user tweets in \qanda. 
We measure the prevalence --  the empirical probability of observing a CDS n-gram in a tweet given an ideological group. 
Additionally, we utilize $100$ bootstrap samples (i.e., repeated sampling of tweets) to estimate the prevalence distributions.
\cref{ide:subfig:dichotomous-thinking} shows that all non-neutral ideologies exhibit a significantly higher prevalence of dichotomous thinking, with right-leaning higher than left-leaning and far-right higher than right-leaning.
We perform T-Tests (95\%) to compare group means and find all differ significantly from each other.
These findings support prior literature \cite{meyer2020political}, and extend it by showing that the far-right might engender an even greater extent of dichotomous thinking.
Other cognitive distortions' prevalences are summarized in the online appendix \cite{appendix}. \section{Conclusion}
\label{ide:sec:Conclusion}
This work proposes a framework for ideology detection pipelines and quantifies biases introduced by ideological proxies. It tests hypotheses of the psychosocial asymmetries of ideological groups, in the online space. We present an evaluation of ideological proxies; qualitatively, indicating proxies that minimize labor, are transferable across multiple contexts, and are available; and, quantitatively, indicating the representativity and robustness of proxies. 
We find the media proxy advantageous, and a pipeline constructed from it and the lexical lens to be optimal, outperforming state-of-the-art approaches. 
Such research is essential for furnishing practitioners with actionable guidelines for ideology detection and its practical applications.

\sh{Limitations.}
The media proxy has several limitations. 
Firstly, it relies on the availability of up-to-date media slant data. 
Publication slant can shift over time, and publication emergence, acquisition and closure can hold significance (particularly on ideological fringes). 
Secondly, some users share media to refute it. 
Thirdly, article slants may differ from publication slants. 
Finally, it will not produce a perfectly representative user subset, although media sharing ubiquity makes it relatively competitive.
Furthermore, our conceptualization of ideology is simplistic, and some political systems are complex requiring complex ideological proxies (which are largely unavailable).

\sh{Future Work.}
We limit our scope to English-speaking Anglo-centric countries due to the expertise and language proficiency of the author team.
However, the study could be applied broadly. 
\citet{newman2021reuters} provides data annually for $46$ diverse countries, including segments of the Global South. 
Our study could be extended to any other uniaxial political setting with little amendment.

{\fontsize{9.0pt}{10.0pt} \selectfont

 }

\section*{Ethics Checklist}

\begin{enumerate}

\item For most authors...
\begin{enumerate}
    \item  Would answering this research question advance science without violating social contracts, such as violating privacy norms, perpetuating unfair profiling, exacerbating the socio-economic divide, or implying disrespect to societies or cultures?
    \answerYes{Yes, see the Ethics and Broader Impact Statement at the end of this checklist.}
  \item Do your main claims in the abstract and introduction accurately reflect the paper's contributions and scope?
    \answerYes{Yes.}
   \item Do you clarify how the proposed methodological approach is appropriate for the claims made? 
    \answerYes{Yes.}
   \item Do you clarify what are possible artifacts in the data used, given population-specific distributions?
    \answerYes{Yes, see Contexts, Datasets, and Ideology Labels.}
  \item Did you describe the limitations of your work?
    \answerYes{Yes, see the Conclusion.}
  \item Did you discuss any potential negative societal impacts of your work?
  \answerYes{Yes, see the Ethics and Broader Impact Statement at the end of this checklist.}
\item Did you discuss any potential misuse of your work?
    \answerYes{Yes, see the Ethics and Broader Impact Statement at the end of this checklist.}
    \item Did you describe steps taken to prevent or mitigate potential negative outcomes of the research, such as data and model documentation, data anonymization, responsible release, access control, and the reproducibility of findings?
    \answerYes{Yes.}
  \item Have you read the ethics review guidelines and ensured that your paper conforms to them?
    \answerYes{Yes.}
\end{enumerate}

\item Additionally, if your study involves hypotheses testing...
\begin{enumerate}
  \item Did you clearly state the assumptions underlying all theoretical results?
    \answerYes{Yes.}
  \item Have you provided justifications for all theoretical results?
  \answerYes{Yes.}
  \item Did you discuss competing hypotheses or theories that might challenge or complement your theoretical results?
  \answerYes{Yes.}
  \item Have you considered alternative mechanisms or explanations that might account for the same outcomes observed in your study?
  \answerYes{Yes.}
  \item Did you address potential biases or limitations in your theoretical framework?
  \answerYes{Yes.}
  \item Have you related your theoretical results to the existing literature in social science?
  \answerYes{Yes.}
  \item Did you discuss the implications of your theoretical results for policy, practice, or further research in the social science domain?
  \answerYes{Yes.}
\end{enumerate}

\item Additionally, if you are including theoretical proofs...
\begin{enumerate}
  \item Did you state the full set of assumptions of all theoretical results?
    \answerNA{NA}
	\item Did you include complete proofs of all theoretical results?
    \answerNA{NA}
\end{enumerate}

\item Additionally, if you ran machine learning experiments...
\begin{enumerate}
  \item Did you include the code, data, and instructions needed to reproduce the main experimental results (either in the supplemental material or as a URL)?
    \answerNo{No, we will include our git repository once the paper gets accepted. The repository will include the code and instructions.}
  \item Did you specify all the training details (e.g., data splits, hyperparameters, how they were chosen)?
    \answerYes{Yes.}
     \item Did you report error bars (e.g., with respect to the random seed after running experiments multiple times)?
    \answerYes{Yes, where applicable.}
	\item Did you include the total amount of compute and the type of resources used (e.g., type of GPUs, internal cluster, or cloud provider)?
    \answerNo{No, our study does not require significant compute resources.}
     \item Do you justify how the proposed evaluation is sufficient and appropriate to the claims made? 
    \answerYes{Yes.}
     \item Do you discuss what is ``the cost`` of misclassification and fault (in)tolerance?
    \answerYes{Yes.}
\end{enumerate}

\item Additionally, if you are using existing assets (e.g., code, data, models) or curating/releasing new assets, \textbf{without compromising anonymity}...
\begin{enumerate}
  \item If your work uses existing assets, did you cite the creators?
    \answerYes{Yes.}
  \item Did you mention the license of the assets?
    \answerNA{NA}
  \item Did you include any new assets in the supplemental material or as a URL?
    \answerYes{Yes.}
  \item Did you discuss whether and how consent was obtained from people whose data you're using/curating?
    \answerNA{NA.}
  \item Did you discuss whether the data you are using/curating contains personally identifiable information or offensive content?
    \answerNA{NA}
\item If you are curating or releasing new datasets, did you discuss how you intend to make your datasets FAIR?
\answerNA{NA}
\item If you are curating or releasing new datasets, did you create a Datasheet for the Dataset? 
\answerNA{NA}
\end{enumerate}

\item Additionally, if you used crowdsourcing or conducted research with human subjects, \textbf{without compromising anonymity}...
\begin{enumerate}
  \item Did you include the full text of instructions given to participants and screenshots?
    \answerNA{NA}
  \item Did you describe any potential participant risks, with mentions of Institutional Review Board (IRB) approvals?
    \answerNA{NA}
  \item Did you include the estimated hourly wage paid to participants and the total amount spent on participant compensation?
    \answerNA{NA}
   \item Did you discuss how data is stored, shared, and deidentified?
   \answerNA{NA}
\end{enumerate}

\end{enumerate}

\section*{Ethical Statement}
This work introduces a powerful tool for inferring user ideology based on covert cues such as language patterns.
We demonstrate our tool for detecting far-right ideologies; 
however, it could, in theory, be used by oppressive regimes to infer the true ideologies of their citizens and expose their opponents~\citep{Radsch2016}.
The Countering Violent Extremism (CVE) literature~\citep{Betz2016} explores the ethical concerns of developing tools that can be used for oppressive ends and proposes mitigation strategies.
There are also privacy concerns, as one's ideology can be viewed as an intimate and private trait that our tool can expose.
Additionally, we show how to use our pipeline to profile entire online populations based on their psychosocial characteristics.
We argue that the pipeline predictions are not prescriptive;
they should be treated as an early warning system, requiring human expert investigation.
We further note that we only use expert-inferred political affiliation as our ground truth, not private self-reported political indicator data.  

\clearpage
\appendix

\section*{Appendices}
This document accompanies the submission. The information in this document complements the submission and is presented here for completeness reasons. It is not required to understand the main paper or reproduce the results.

\section{Dataset Collection Details}
\sh{\qanda.} 
We collect discussions related to the Australian panel show Q+A~\citep{Q+Ashow2023}, where panelists (public figures, politicians, and experts) answer curated audience questions.
Twitter participation is encouraged in airings. 
We collect \qanda using the filter keyword \emph{qanda} during January-December 2020. 

\sh{\ausvotes.} 
We collect discussions about the 2022 Australian Federal election, tracking the lead-up and aftermath. 
It follows the major parties and their leaders: the left-leaning Australian Labor Party led by Anthony Albanese and the right-leaning Liberal-National Coalition led by Scott Morrison. 
We collect \ausvotes using the keywords \emph{auspol} and \emph{ausvotes}, and for mentions of \emph{@ScottMorrisonMP}, \emph{@AlboMP}, and \emph{@AusElectoralCom}, between 9 May and 15 June 2022 (the elections occurred on 21 May).

\sh{\socialsense} \citep{calderon2022opinion} features discussions related to the Australian Black Summer bushfires, which gathered discourse concerning climate change, and contains far-right opinions. 
\socialsense contains $90$ days of Twitter and Facebook discussions, from 1 November to 29 January 2020.

\sh{\riot} \cite{riot} features discussions about the January 6th US Capitol Insurrection, including election fraud and insurrection topics. 
The dataset spans 6 January to 1 February 2021 and was collected with the filter keywords \emph{TrumpRally}, \emph{Democracy}, \emph{USCapitol}, \emph{Capitol}, \emph{DCProtests}, and \emph{AshliBabbit}.

\sh{\parler{}} \cite{parler} features discussions about the US Capitol Insurrection from Parler.
We collect all posts emitted during the day of 6 January 2021.

\section{All UUS/UUS+ Metrics}
Thi section shows all possible runs for the \emph{UUS} and \emph{UUS+}. We notice that in many instances \emph{UUS} fails to seperate clusters, and even in instances where seperation can be achieved many suffer from poor performance. This shows that these techniques lack robustness for more difficult datasets.
\begin{table}[h]
    \caption{
        \textbf{All Baseline Performances.} The table shows to performances for all combinations of the \emph{UUS} and \emph{UUS+} baselines.
      }
    \label{ide:tab:all_baseline_performances}
    \begin{tabularx}{\columnwidth}{lX|XXX}
    \hline
    \textbf{Representation} & \textbf{Active Users} & \textbf{F1-Macro} & \textbf{AUC ROC} & \textbf{UUS F1-Macro} \\ \hline
    H                       & 500                   & 0.37              & 0.68             & 0.37                  \\
    H                       & 1000                  & -                 & -                & -                     \\
    H                       & 5000                  & 0.37              & 0.54             & 0.37                  \\
    HR                      & 500                   & -                 & -                & -                     \\
    HR                      & 1000                  & -                 & -                & -                     \\
    HR                      & 5000                  & 0.89              & 0.92             & 0.85                  \\
    R                       & 500                   & -                 & -                & -                     \\
    R                       & 1000                  & -                 & -                & -                     \\
    R                       & 5000                  & 0.93              & 0.93             & 0.87                  \\
    T                       & 500                   & -                 & -                & -                     \\
    T                       & 1000                  & -                 & -                & -                     \\
    T                       & 5000                  & -                 & -                & -                     \\
    TH                      & 500                   & 0.4               & 0.58             & 0.54                  \\
    TH                      & 1000                  & -                 & -                & -                     \\
    TH                      & 5000                  & 0.92              & 0.91             & 0.87                  \\
    TR                      & 500                   & -                 & -                & -                     \\
    TR                      & 1000                  & -                 & -                & -                     \\
    TR                      & 5000                  & -                 & -                & -                     \\
    TRH                     & 500                   & -                 & -                & -                     \\
    TRH                     & 1000                  & -                 & -                & -                     \\
    TRH                     & 5000                  & 0.41              & 0.75             & 0.36                 
    \end{tabularx}
\end{table}

\section{Left-Right Annotation Procedure}
Ideology is the subject of considerable subjectivity, not only because experts have their own ideology, but because annotators are often unclear as to what evidence is permissible for use. 
For this task we issued the following guidelines to annotators:

\begin{quote}
It is not always clear what should count as an ideological signal. For our purposes, we will include the following as signals of ideology:
\begin{itemize}
    \item If a target user promotes/retweets someone or an organisation with a known ideological affiliation, you may assume that the target endorse them. For example, if a target user retweets a labor MP then you can label the user as 'left'.
    \item If a target user, has a stance against someone with a known ideological affiliation, then you might infer that the target user's ideology is the opposing ideology. For example, if a target user calls a labor MP an insult, then you can label the user as 'right'.
    \item If a target user expresses a view about a issue related to an ideology, you can infer the user's ideology. For example, if a user supports LGBTQ or environmental issues, then (if there is enough evidence) you may label them as 'left'.
\end{itemize}
\end{quote}

These guidelines aim to increase the clarity of the annotation task. In countries where political affiliation is obvert (e.g. the united states), this labelling task is often unambiguous; however, in Australia ideological signals are often implicit.
The full annotation briefing material is available in the code repository [\url{https://github.com/behavioral-ds/ideology_prediction}].

\section*{Context-Transfer Illustration}
\begin{figure}[tbp]
    \newcommand\myheight{0.29}
    \centering
\includegraphics[width=\columnwidth]{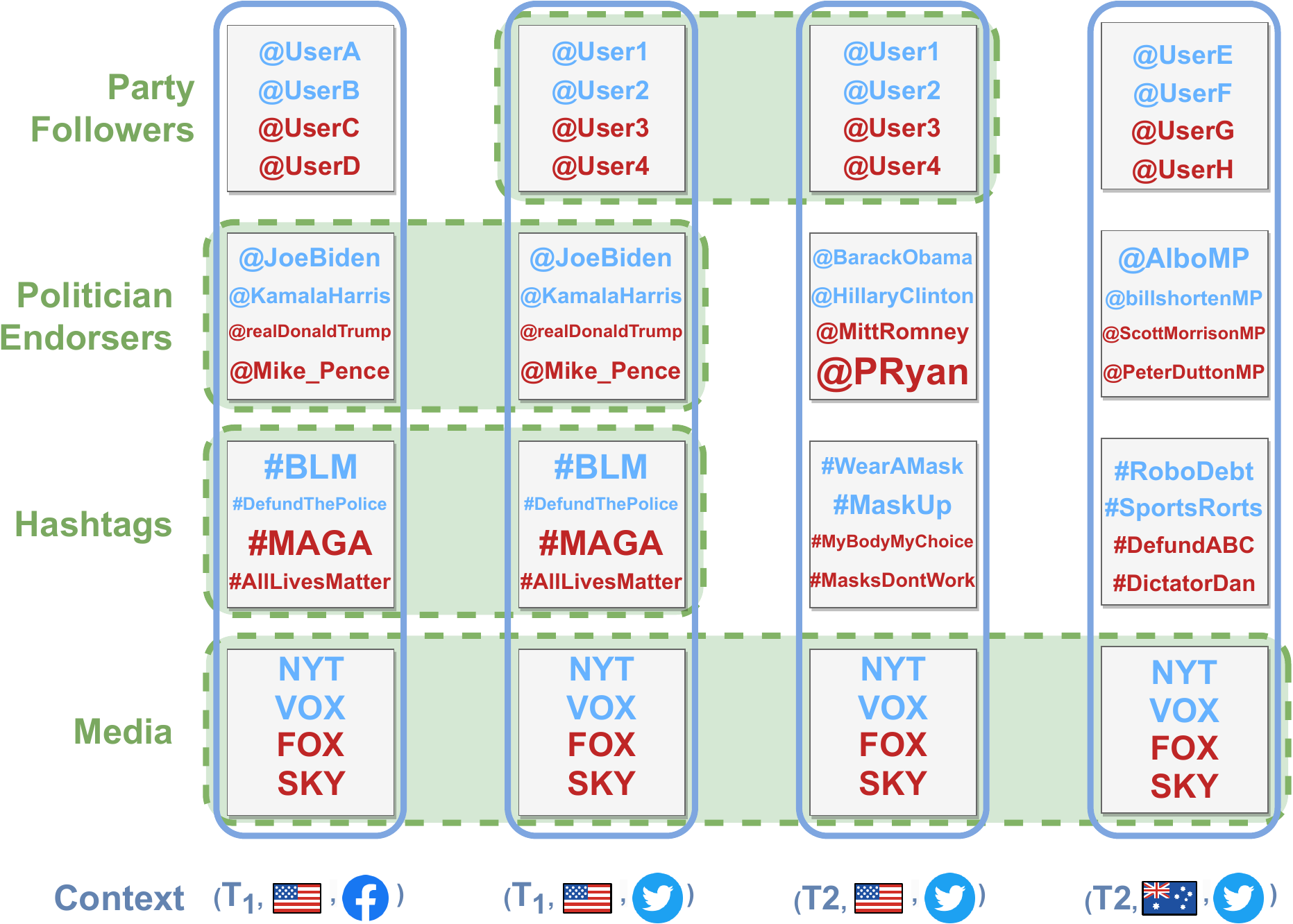}
\caption{
        \textbf{Most ideology proxies do not generalize across contexts.}    
        The x-axis shows four contexts that vary in time ($T_1$ and $T_2$), country (Australia and USA), and platform (Twitter and Facebook).
        The y-axis show four proxies: endorsing political parties or political figures, using politically charged hashtags and the consumed media slant.
        The {\color{olive}{green dashed boxes}} indicate whether a proxy is applicable across contexts. 
}
    \label{ide:fig:teaser}
\end{figure} \cref{ide:fig:teaser} further illustrates the difficulty with utilizing particular proxies as ground truth. 
We observe that some ideological proxies are consistent across only some contexts (represented by the dashed green boxes).
For example, \texttt{\#RoboDebt} (in relation to an Australian incident) is not relevant to the USA and did not exist before 2016; and, although \texttt{@MittRomney} signaled right-wing ideology in 2012, the right has shifted since Trump's election.

Prior ideology detection techniques fail to easily \emph{context-switch} and cannot be readily applied to multiple distinct domains.

\section*{Media Publication Slants}
\begin{figure*}[]
    \newcommand\myheight{0.25}
    \centering
    \includegraphics[width=\textwidth]{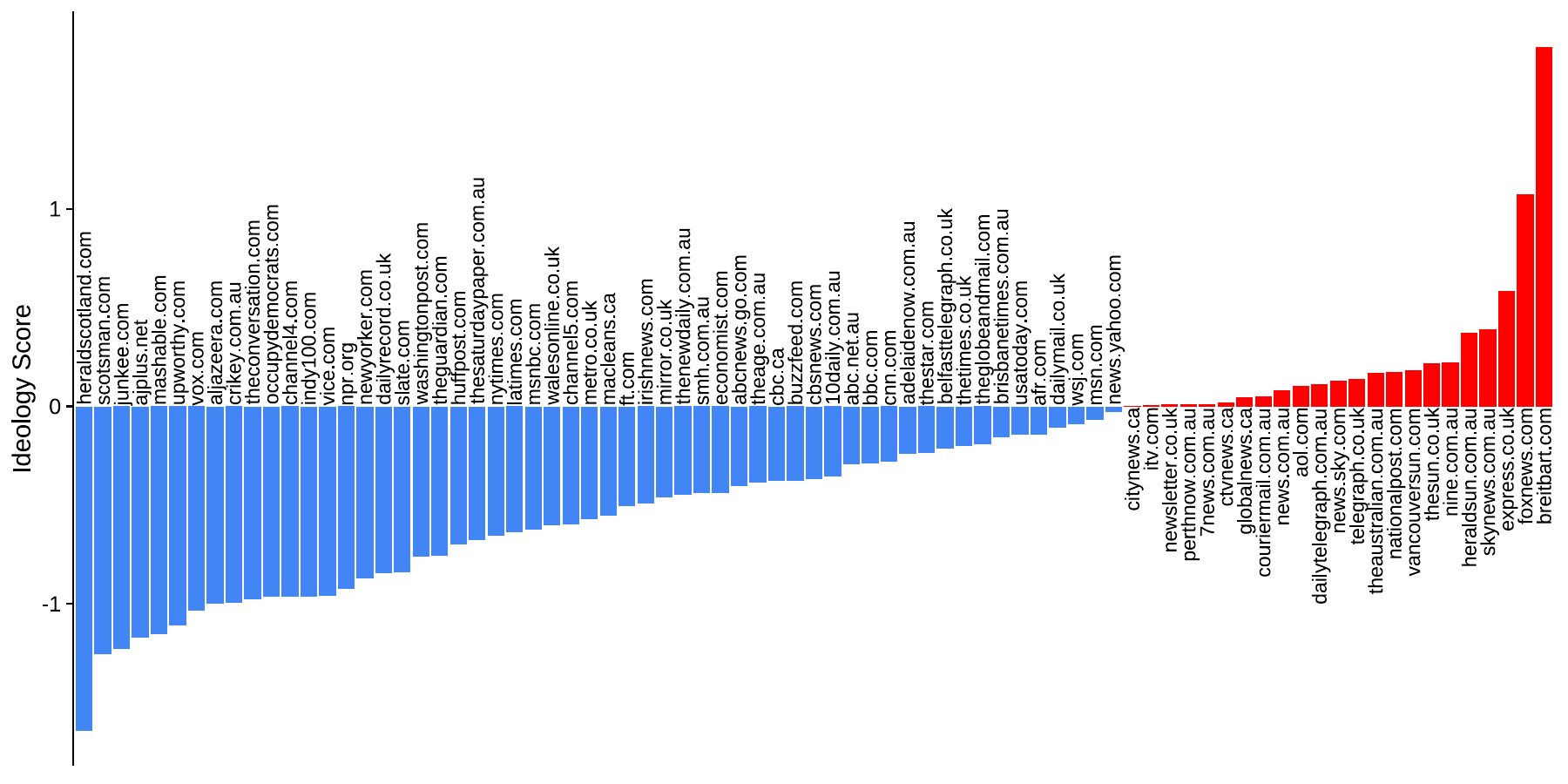}\caption{
        \textbf{Media Publication Slants.} The plot shows the slants of Media Publications, as averaged over the year, country, and source point estimates.
    }
    \label{ide:fig:media_slants}
\end{figure*}

The media slant scores are shown in \cref{ide:fig:media_slants}, where we observed publications like \emph{Breitbart} and \emph{Fox News} are extremely right-leaning, and \emph{Vox} and \emph{NYTimes} are left-leaning.

\section*{Cognitive Distortions Schemata Prevalence}
\begin{figure*}[tbp]
    \centering
    \includegraphics[width=\textwidth]{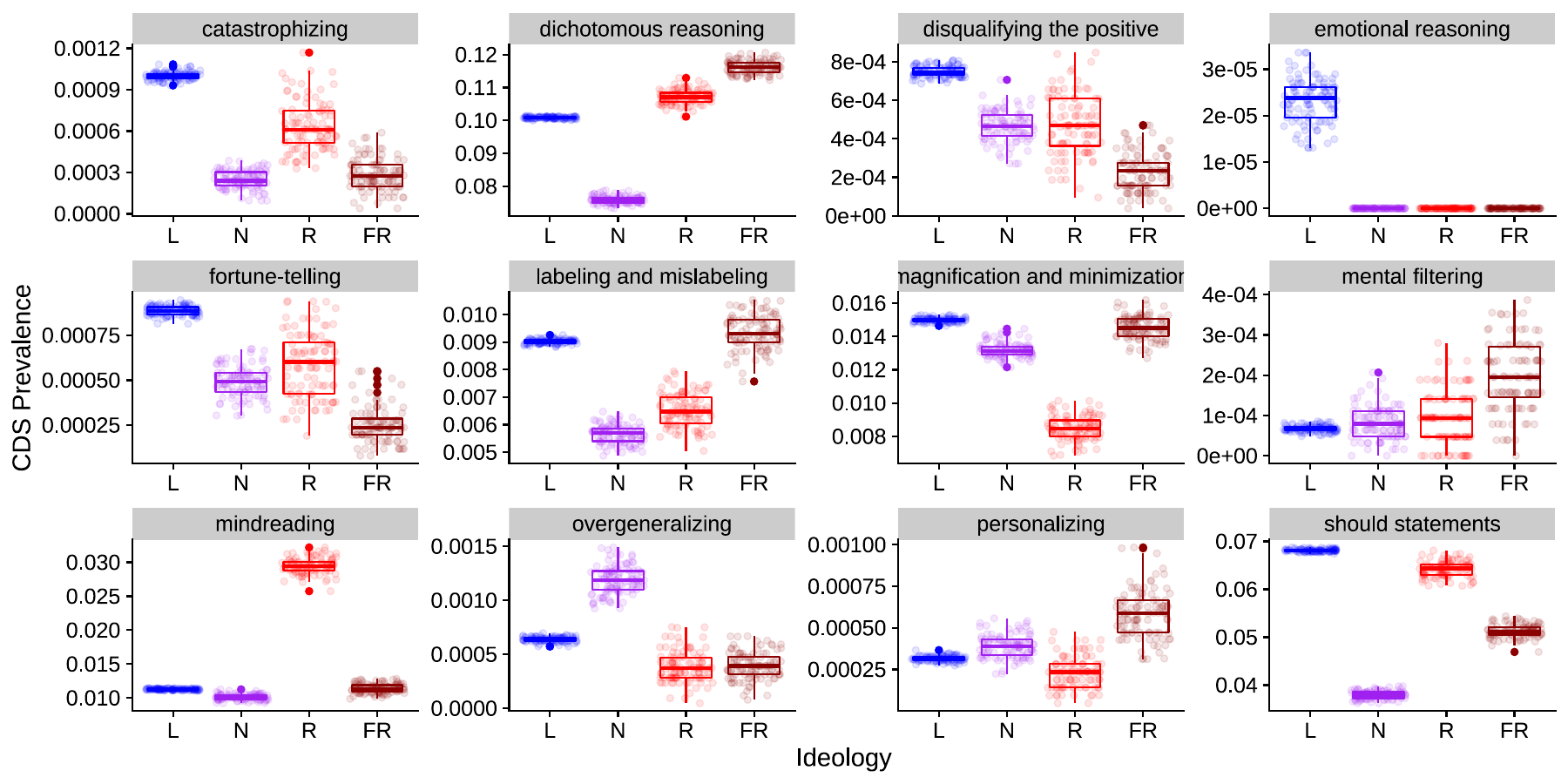}\caption{CDS Prevalence}
    \label{ide:fig:cds_boxplots}
\end{figure*}

\cref{ide:fig:cds_boxplots} shows the prevalence of all twelve cognitive distortions in each of the ideological groups, for \qanda. Note that many CDS n-grams are extremely rare (or do not appear), namely; \emph{emotional reasoning} and \emph{mental filtering}. In several CDS the left exhibit higher prevalence, such as \emph{catastrophizing}, \emph{fortune-telling}, \emph{disqualifying the positive}, and \emph{should statements}.

\section*{Flag Emoji Hurdle Model}
\begin{figure}[h]
    \newcommand\myheight{0.20}
    \centering
    \includegraphics[width=\columnwidth]{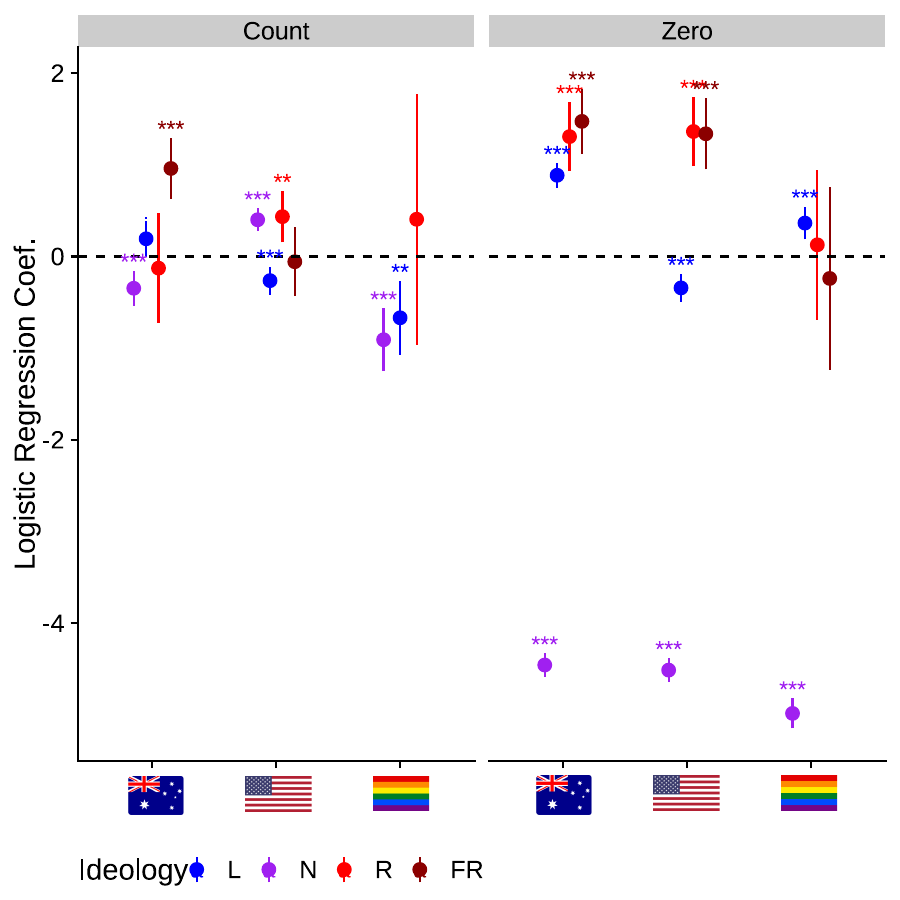}\caption{Hurdle Model}
    \label{ide:fig:hurdle_model}
\end{figure}

For completeness, we present the results of the hurdle model (used to model zero-inflated count data, such a tokens in a corpus). The hurdle model is a mixed model, comprised of a logistic regression to model the presense of no emoji, and a truncated poisson with log linkage, to model the count of the emoji. \cref{ide:fig:hurdle_model} shows the coefficients for each model, including the reference groups. 
\emojiAUS is observed more for far-right users in both the zero and the count models. The count models for the other flags show mixed results and not significant.

\section*{Precision-Recall of Pipelines}
\begin{figure}[]
    \newcommand\myheight{0.33}
    \centering
    \includegraphics[width=\columnwidth]{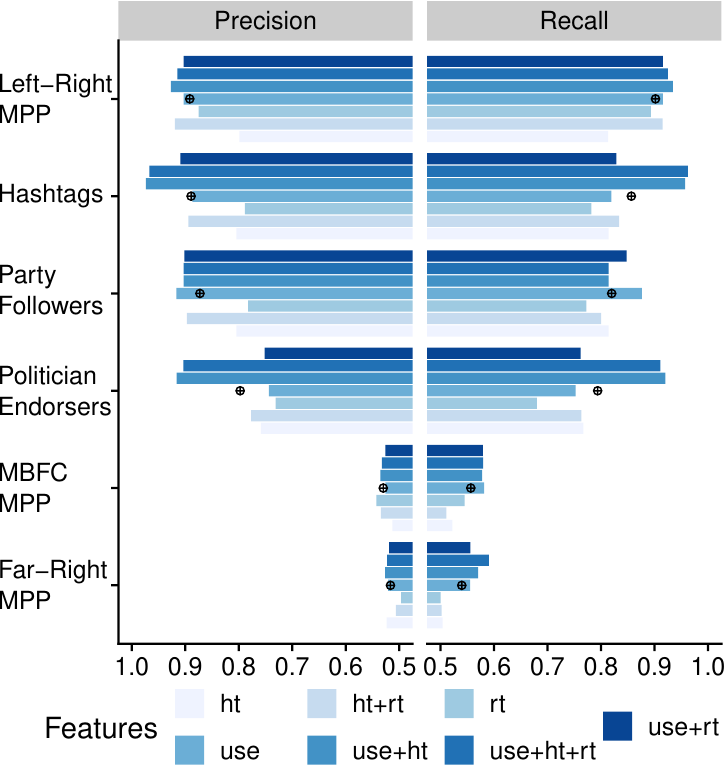}\caption{
        \textbf{Precision-Recall.} The plot shows the macro-averaged precision and recall of pipelines, trained with each proxy (y-axis) and each feature set (colors), probability calibrated with the hold-out validation set for F1-macro scores.
    }
    \label{ide:fig:precision_recall_plot}
\end{figure}

\cref{ide:fig:precision_recall_plot} shows precision and recall for every lens combinations and proxy.

\section*{Predicted Label Distribution}
\cref{ide:tab:label_distribution} shows the distribution of predicted labels, to provide context for the psychosocial analysis.
\begin{table}[tb]
    \centering
    \caption{
        \textbf{Distribution of Predicted Labels.} 
        The number of users predicted to be in each class (rows) for each dataset (columns).
        Note that for many datasets there is a significant imbalance toward the left (except \parler which is a right-leaning platform).
}
    \label{ide:tab:label_distribution}
\begin{tabularx}{\columnwidth}{l|XXXXX}
        \toprule
        & \rohhead{\qanda} & \rohhead{\ausvotes} & \rohhead{\socialsense} & \rohhead{Riot} & \rohhead{Parler}\\
        \midrule
        Left & 80,375 & 189,233 & 48,056 & 339,095 & 293\\
        Neutral & 21,176 & 79,221 & 604 & 227,839 & 48,829\\
        Right & 777 & 3,689 & 464 & 3,624 & 68,104\\
        Far-right & 746 & 1,731 & 318 & 3,723 & 2,822\\
        \bottomrule
    \end{tabularx}
\end{table}
     
\section*{Dataset Profiling}
\begin{figure}[]
    \newcommand\myheight{0.25}
    \centering
    \includegraphics[width=\columnwidth]{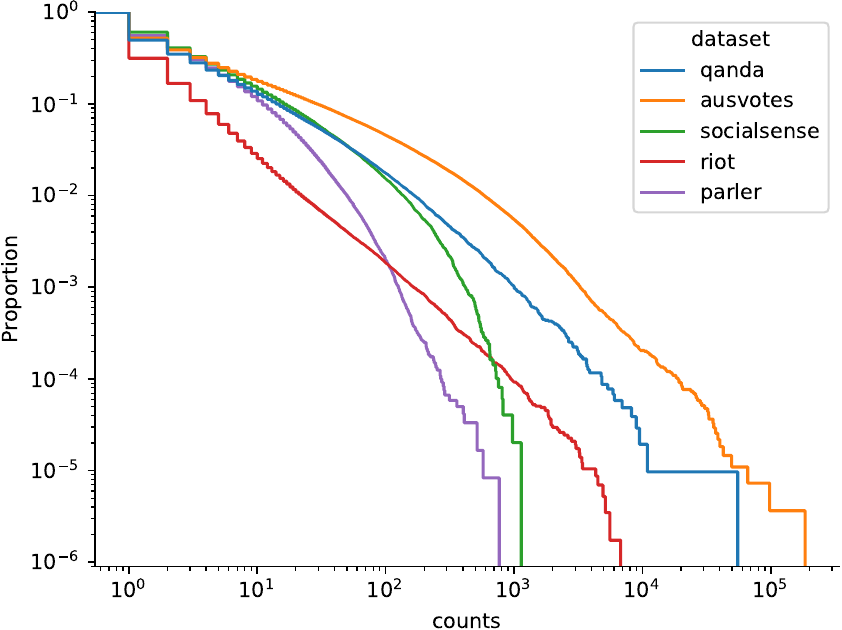}\caption{
        \textbf{Activity Distribution.} The log-log ECCDF distribution of activity (number of posts per user) for each dataset. 
    }
    \label{ide:fig:activity_plot}
\end{figure}

Activity levels are often a concern for ideology detection frameworks, given that low-activity users reveal few signals of ideology.
\cref{ide:fig:activity_plot} shows the distribution of activity for users for each dataset. It shows long-tailed activity distributions and the proportion of low-activity users. \riot shows a significant proportion of low-activity users, who're often difficult to classify.

\section*{Exhaustive Psychosocial Analysis}
\subsection*{Grievance}
This section shows the difference between ideological groups in terms of grievance categories for all available datasets.
\begin{figure}[h]
    \newcommand\myheight{0.20}
    \centering
    \includegraphics[height=\myheight\textheight]{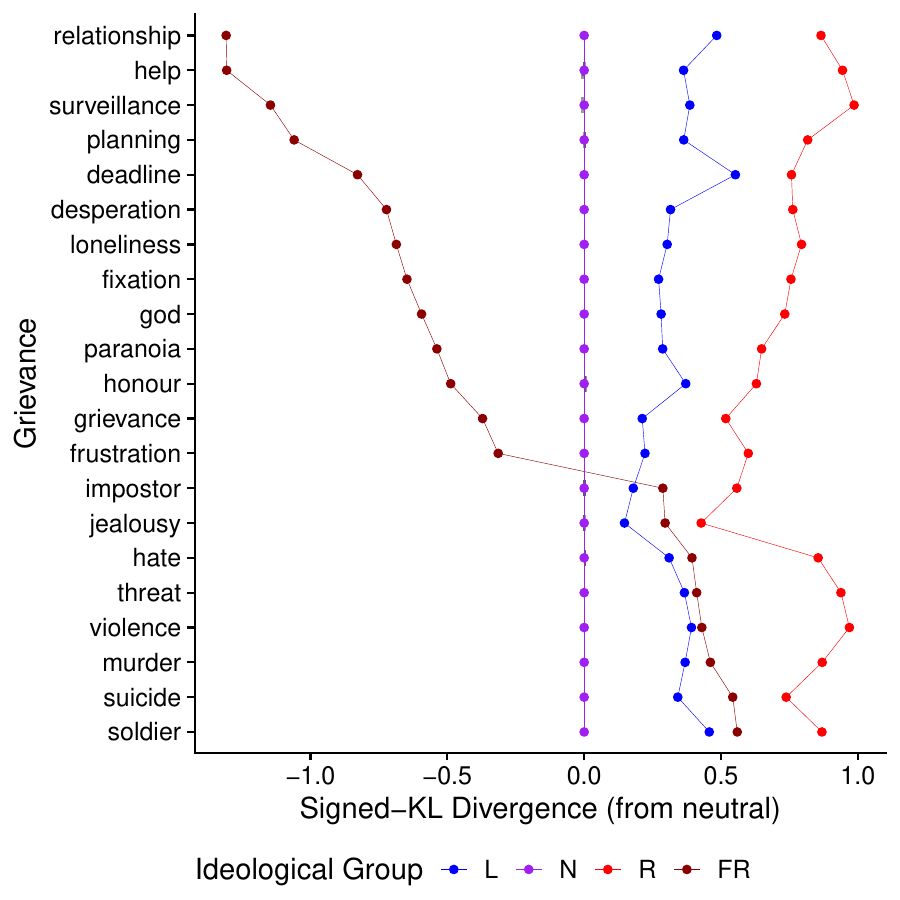}\caption{Grievance \qanda}
\end{figure}

\begin{figure}[h]
    \newcommand\myheight{0.20}
    \centering
    \includegraphics[height=\myheight\textheight]{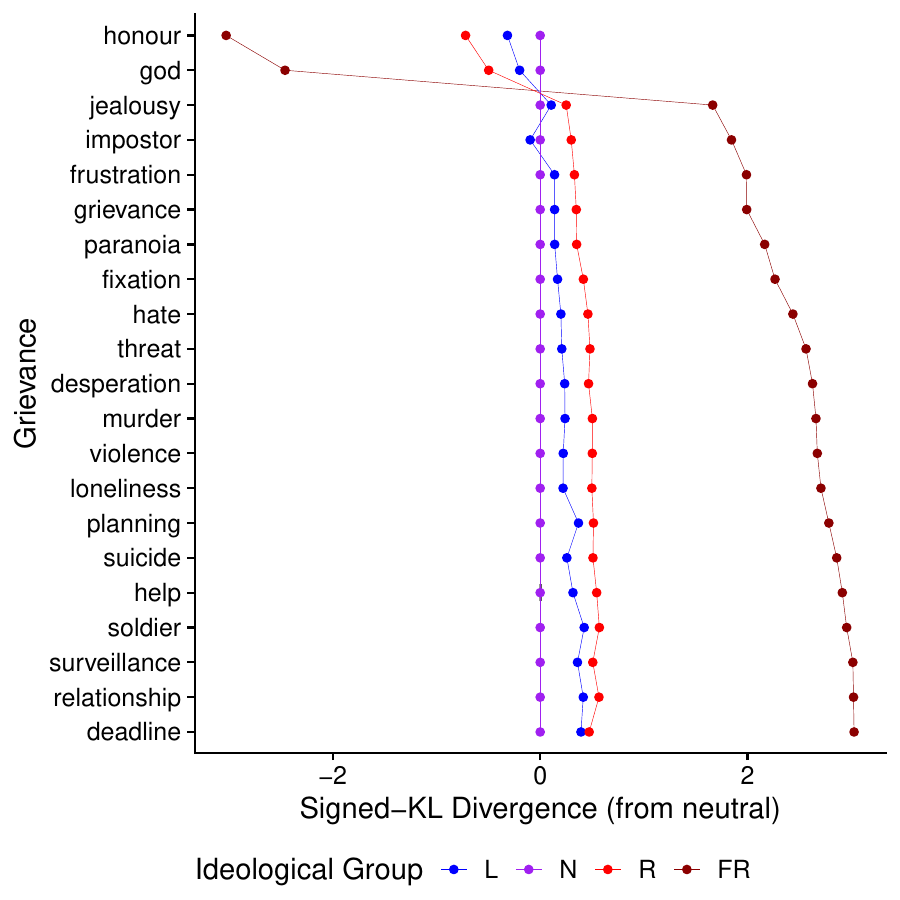}\caption{Grievance \ausvotes}
\end{figure}

\begin{figure}[h]
    \newcommand\myheight{0.20}
    \centering
    \includegraphics[height=\myheight\textheight]{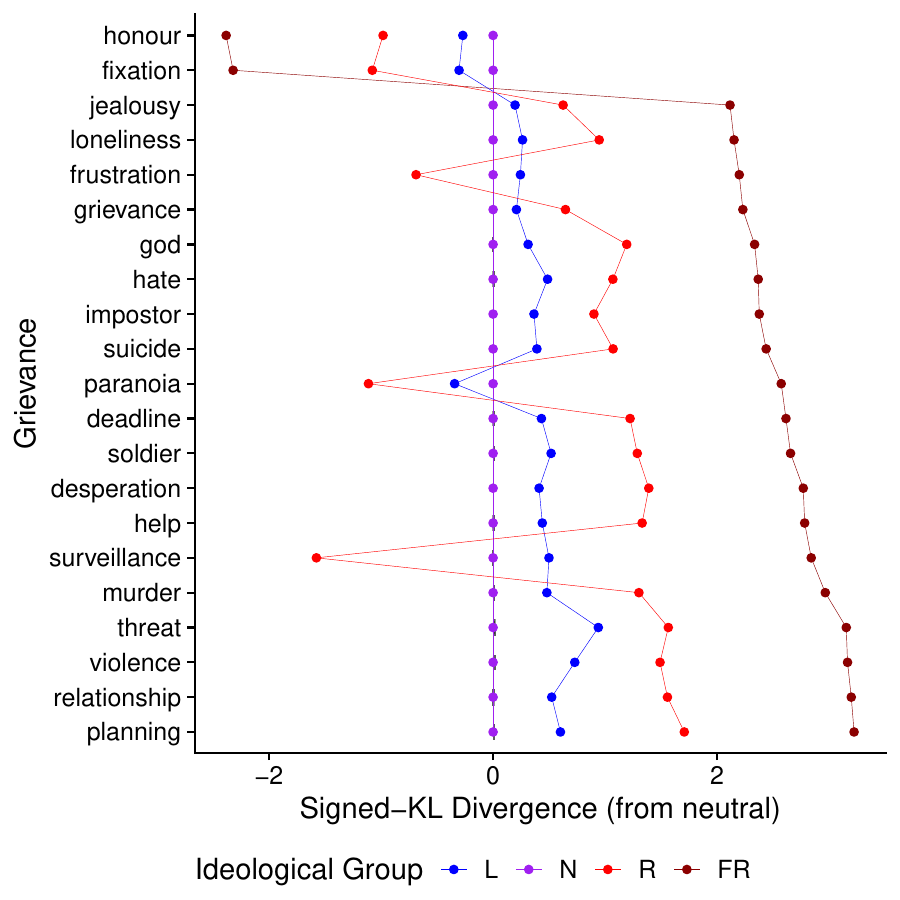}\caption{Grievance \socialsense}
\end{figure}

\begin{figure}[h]
    \newcommand\myheight{0.20}
    \centering
    \includegraphics[height=\myheight\textheight]{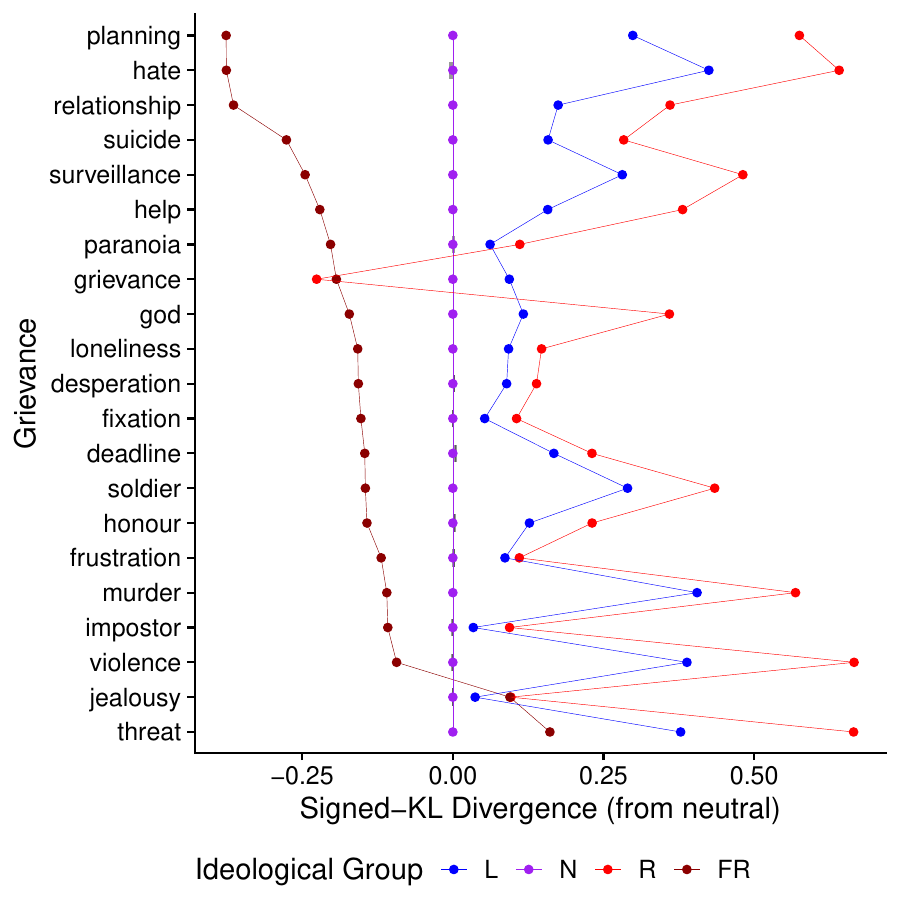}\caption{Grievance \riot}
\end{figure}

\begin{figure}[h]
    \newcommand\myheight{0.20}
    \centering
    \includegraphics[height=\myheight\textheight]{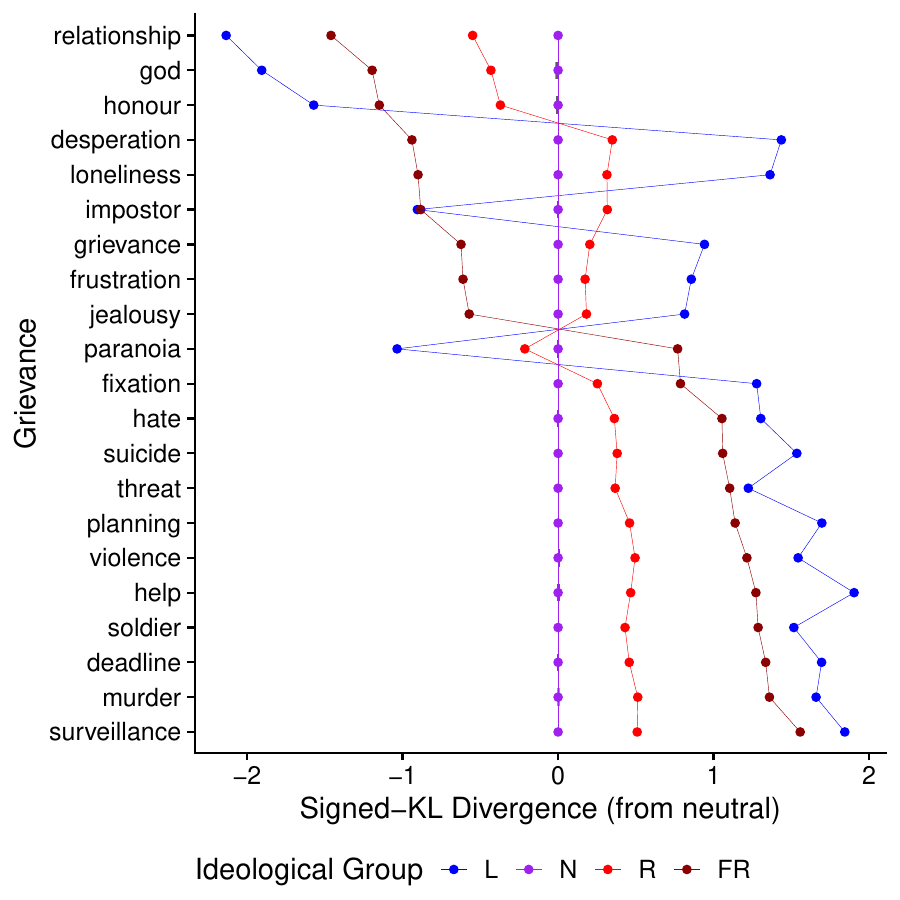}\caption{Grievance \parler}
\end{figure}

\subsection*{MFT}
This section shows the difference between ideological groups in terms of moral foundations for all available datasets.
\begin{figure}[h]
    \newcommand\myheight{0.20}
    \centering
    \includegraphics[height=\myheight\textheight]{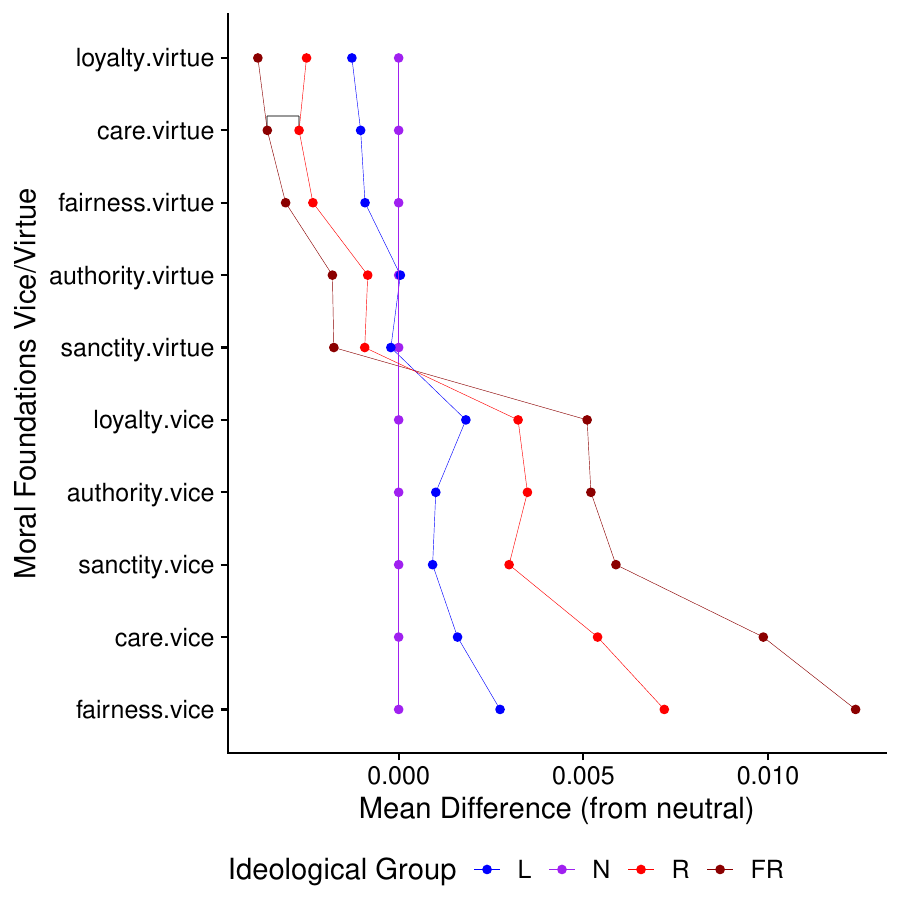}\caption{MFT \qanda}
\end{figure}

\begin{figure}[h]
    \newcommand\myheight{0.20}
    \centering
    \includegraphics[height=\myheight\textheight]{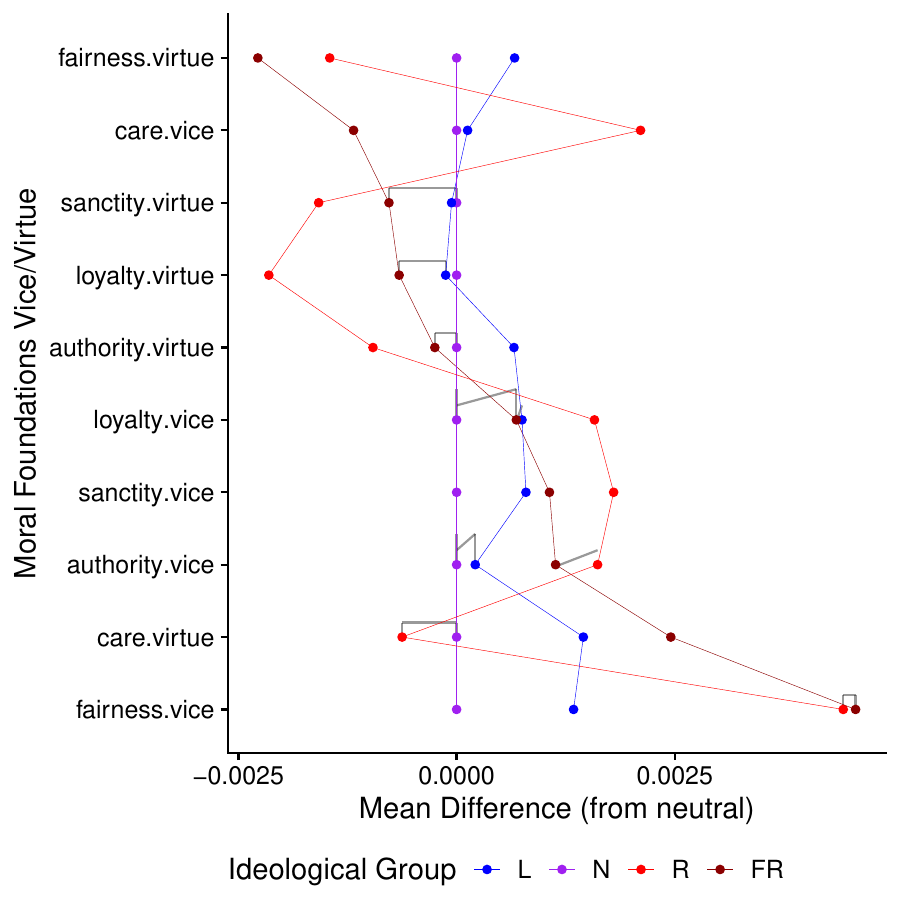}\caption{MFT \ausvotes}
\end{figure}

\begin{figure}[h]
    \newcommand\myheight{0.20}
    \centering
    \includegraphics[height=\myheight\textheight]{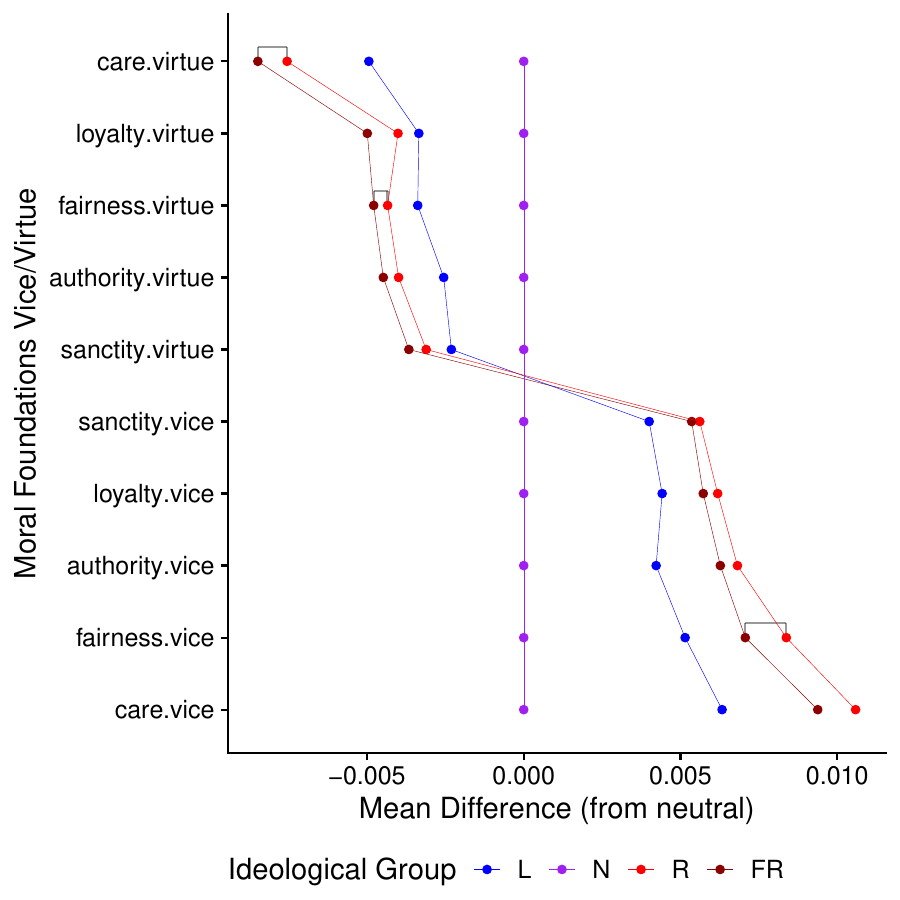}\caption{MFT \socialsense}
\end{figure}

\begin{figure}[h]
    \newcommand\myheight{0.20}
    \centering
    \includegraphics[height=\myheight\textheight]{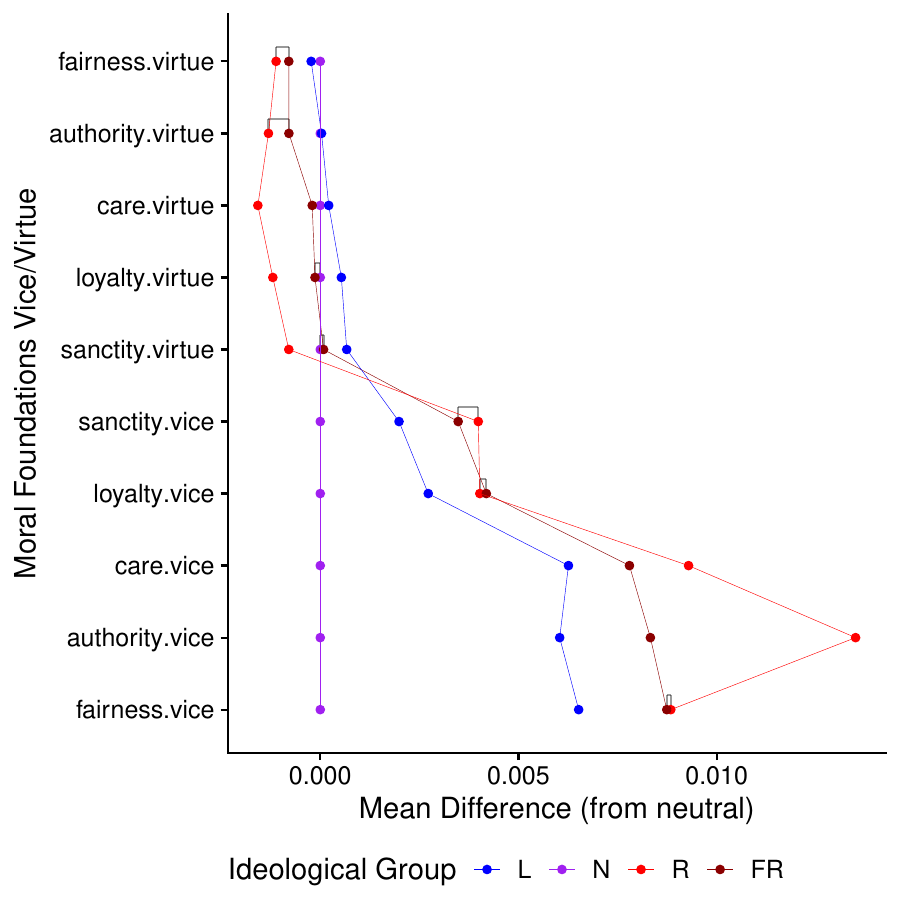}\caption{MFT \riot}
\end{figure}

\begin{figure}[h]
    \newcommand\myheight{0.20}
    \centering
    \includegraphics[height=\myheight\textheight]{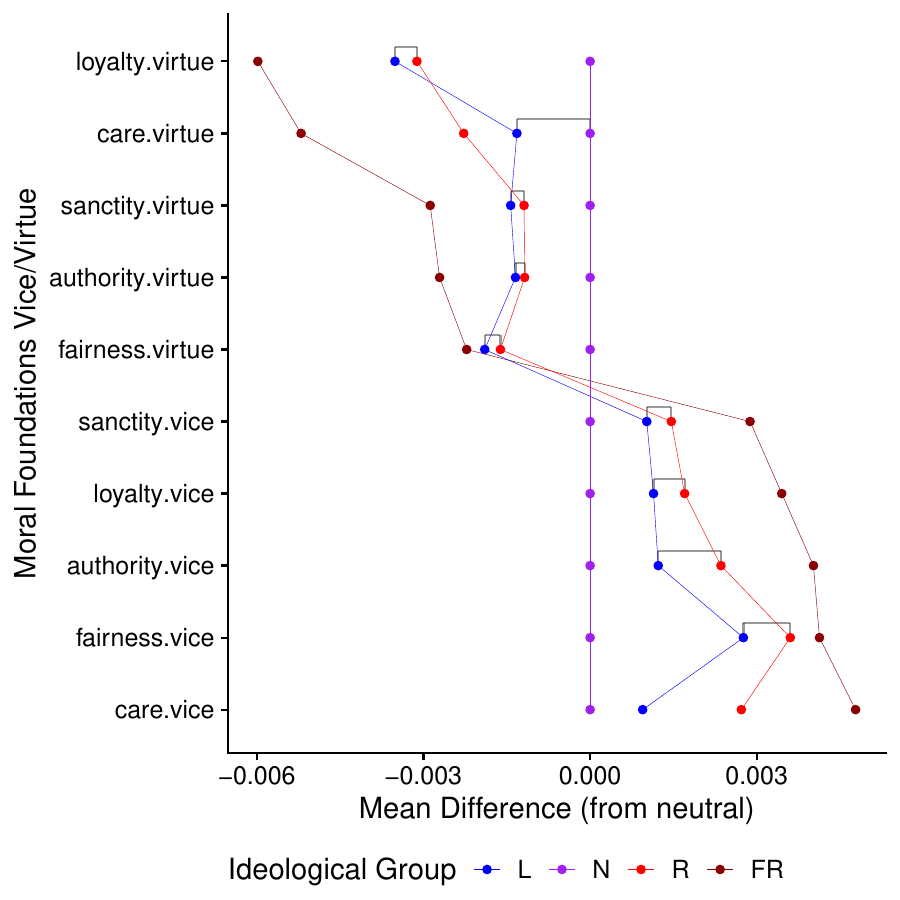}\caption{MFT \parler}
\end{figure} 

\begin{thebibliography}{75}
\providecommand{\natexlab}[1]{#1}

\bibitem[{Aldayel and Magdy(2019)}]{aldayel2019your}
Aldayel, A.; and Magdy, W. 2019.
\newblock Your stance is exposed! analysing possible factors for stance
  detection on social media.
\newblock \emph{CSCW}.

\bibitem[{Aliapoulios et~al.(2021)Aliapoulios, Bevensee, Blackburn, Bradlyn,
  Cristofaro, Stringhini, and Zannettou}]{parler}
Aliapoulios, M.; Bevensee, E.; Blackburn, J.; Bradlyn, B.; Cristofaro, E.~D.;
  Stringhini, G.; and Zannettou, S. 2021.
\newblock {A Large Open Dataset from the Parler Social Network}.

\bibitem[{Alizadeh et~al.(2019)Alizadeh, Weber, Cioffi-Revilla, Fortunato, and
  Macy}]{alizadeh2019psychology}
Alizadeh, M.; Weber, I.; Cioffi-Revilla, C.; Fortunato, S.; and Macy, M. 2019.
\newblock Psychology and morality of political extremists: evidence from
  Twitter language analysis of alt-right and Antifa.
\newblock \emph{EPJ DS}.

\bibitem[{Alkiek, Zhang, and Jurgens(2022)}]{alkiek2022classification}
Alkiek, K.; Zhang, B.; and Jurgens, D. 2022.
\newblock Classification without (Proper) Representation: Political
  Heterogeneity in Social Media and Its Implications for Classification and
  Behavioral Analysis.
\newblock In \emph{ACL}.

\bibitem[{AllSides(2022)}]{allsides}
AllSides. 2022.
\newblock AllSides Media Bias Ratings.
\newblock \url{https://www.allsides.com/media-bias/ratings}.
\newblock Accessed: 2022-04-08.

\bibitem[{An, Quercia, and Crowcroft(2014)}]{an2014partisan}
An, J.; Quercia, D.; and Crowcroft, J. 2014.
\newblock Partisan sharing: Facebook evidence and societal consequences.
\newblock In \emph{COSN}.

\bibitem[{Appendix(2024)}]{appendix}
Appendix, O. 2024.
\newblock Supplementary Material: Practical Guidelines for Ideology Detection
  Pipelines and Psychosocial Applications.
\newblock \url{https://bit.ly/ideology_detection}.

\bibitem[{Auxier and Anderson(2021)}]{auxier2021social}
Auxier, B.; and Anderson, M. 2021.
\newblock Social media use in 2021.
\newblock \emph{Pew Research Center}.

\bibitem[{Badaan et~al.(2023)Badaan, Hoffarth, Roper, Parker, and
  Jost}]{badaan2023ideological}
Badaan, V.; Hoffarth, M.; Roper, C.; Parker, T.; and Jost, J.~T. 2023.
\newblock Ideological asymmetries in online hostility, intimidation, obscenity,
  and prejudice.
\newblock \emph{Scientific reports}.

\bibitem[{Badawy, Lerman, and Ferrara(2019)}]{badawy2019falls}
Badawy, A.; Lerman, K.; and Ferrara, E. 2019.
\newblock Who falls for online political manipulation?
\newblock In \emph{WWW}.

\bibitem[{Bailo, Johns, and Rizoiu(2023)}]{Bailo2023}
Bailo, F.; Johns, A.; and Rizoiu, M.-A. 2023.
\newblock {Riding information crises: the performance of far-right Twitter
  users in Australia during the 2019--2020 bushfires and the COVID-19
  pandemic}.
\newblock \emph{Information, Communication \& Society}.

\bibitem[{Bakshy, Messing, and Adamic(2015)}]{bakshy2015exposure}
Bakshy, E.; Messing, S.; and Adamic, L.~A. 2015.
\newblock Exposure to ideologically diverse news and opinion on Facebook.
\newblock \emph{Science}.

\bibitem[{Barber{\'a}(2015)}]{barbera2015birds}
Barber{\'a}, P. 2015.
\newblock Birds of the same feather tweet together: Bayesian ideal point
  estimation using Twitter data.
\newblock \emph{Political analysis}.

\bibitem[{Bathina et~al.(2021)Bathina, Ten~Thij, Lorenzo-Luaces, Rutter, and
  Bollen}]{bathina2021individuals}
Bathina, K.~C.; Ten~Thij, M.; Lorenzo-Luaces, L.; Rutter, L.~A.; and Bollen, J.
  2021.
\newblock Individuals with depression express more distorted thinking on social
  media.
\newblock \emph{Nature Human Behaviour}.

\bibitem[{Betz(2016)}]{Betz2016}
Betz, M. 2016.
\newblock Constraints and opportunities: what role for media development in
  countering violent extremism?

\bibitem[{Bode et~al.(2013)Bode, Hanna, Sayre, Yang, and
  Shah}]{bode2013mapping}
Bode, L.; Hanna, A.; Sayre, B.; Yang, J.; and Shah, D.~V. 2013.
\newblock Mapping the political Twitterverse: Finding connections between
  political elites.

\bibitem[{Booth et~al.(2024)Booth, Lee, Rizoiu, and Farid}]{Booth2024}
Booth, E.; Lee, J.; Rizoiu, M.-A.; and Farid, H. 2024.
\newblock Conspiracy, misinformation, radicalisation: understanding the online
  pathway to indoctrination and opportunities for intervention.
\newblock \emph{Journal of Sociology}.

\bibitem[{Calderon, Ram, and Rizoiu(2024)}]{calderon2022opinion}
Calderon, P.; Ram, R.; and Rizoiu, M.-A. 2024.
\newblock Opinion Market Model: Stemming Far-Right Opinion Spread using
  Positive Interventions.
\newblock In \emph{ICWSM}.

\bibitem[{Cann, Weaver, and Williams(2021)}]{cann2021ideological}
Cann, T.~J.; Weaver, I.~S.; and Williams, H.~T. 2021.
\newblock Ideological biases in social sharing of online information about
  climate change.
\newblock \emph{Plos one}.

\bibitem[{Carr et~al.(2022)Carr, Dancho, Michaud, Chiang, Damoff, Lloyd,
  MacGregor, McKinnon, Noormohamed, Schiefke, Shipley, Popta, and
  Zuberi}]{SAFETY2022}
Carr, H.~J.; Dancho, R.; Michaud, K.; Chiang, P.; Damoff, P.; Lloyd, D.;
  MacGregor, A.; McKinnon, R.; Noormohamed, T.; Schiefke, P.; Shipley, D.;
  Popta, T.~V.; and Zuberi, S. 2022.
\newblock {Rise of Ideologically Motivated Violent Extremism in Canada}.
\newblock Technical report.

\bibitem[{Cer et~al.(2018)Cer, Yang, Kong, Hua, Limtiaco, John, Constant,
  Guajardo-Cespedes, Yuan, Tar et~al.}]{cer2018universal}
Cer, D.; Yang, Y.; Kong, S.-y.; Hua, N.; Limtiaco, N.; John, R.~S.; Constant,
  N.; Guajardo-Cespedes, M.; Yuan, S.; Tar, C.; et~al. 2018.
\newblock Universal sentence encoder.
\newblock \emph{arXiv preprint arXiv:1803.11175}.

\bibitem[{Chakraborty, Goyal, and Mukherjee(2022)}]{chakraborty2022fast}
Chakraborty, S.; Goyal, P.; and Mukherjee, A. 2022.
\newblock Fast Few Shot Self-attentive Semi-supervised Political Inclination
  Prediction.
\newblock In \emph{ICADL}.

\bibitem[{Cichocka et~al.(2016)Cichocka, Bilewicz, Jost, Marrouch, and
  Witkowska}]{cichocka2016grammar}
Cichocka, A.; Bilewicz, M.; Jost, J.~T.; Marrouch, N.; and Witkowska, M. 2016.
\newblock On the grammar of politics—or why conservatives prefer nouns.
\newblock \emph{Political Psychology}.

\bibitem[{Cohen and Ruths(2013)}]{cohen2013classifying}
Cohen, R.; and Ruths, D. 2013.
\newblock Classifying political orientation on Twitter: It’s not easy!
\newblock In \emph{ICWSM}.

\bibitem[{Cohrs(2012)}]{cohrs2012ideological}
Cohrs, J.~C. 2012.
\newblock Ideological bases of violent conflict.

\bibitem[{Darwish et~al.(2020)Darwish, Stefanov, Aupetit, and
  Nakov}]{darwish2020unsupervised}
Darwish, K.; Stefanov, P.; Aupetit, M.; and Nakov, P. 2020.
\newblock Unsupervised user stance detection on twitter.
\newblock In \emph{ICWSM}.

\bibitem[{Eady et~al.(2020)Eady, Bonneau, Tucker, and Nagler}]{eady2020news}
Eady, G.; Bonneau, R.; Tucker, J.~A.; and Nagler, J. 2020.
\newblock News sharing on social media: Mapping the ideology of news media
  content, citizens, and politicians.

\bibitem[{Garimella et~al.(2021)Garimella, Smith, Weiss, and
  West}]{garimella2021political}
Garimella, K.; Smith, T.; Weiss, R.; and West, R. 2021.
\newblock Political polarization in online news consumption.
\newblock In \emph{ICWSM}.

\bibitem[{Graham, Haidt, and Nosek(2009)}]{graham2009liberals}
Graham, J.; Haidt, J.; and Nosek, B.~A. 2009.
\newblock Liberals and conservatives rely on different sets of moral
  foundations.
\newblock \emph{Journal of personality and social psychology}.

\bibitem[{Gu et~al.(2016)Gu, Chen, Sun, and Wang}]{gu2016ideology}
Gu, Y.; Chen, T.; Sun, Y.; and Wang, B. 2016.
\newblock Ideology detection for twitter users with heterogeneous types of
  links.
\newblock \emph{arXiv preprint arXiv:1612.08207}.

\bibitem[{Hopkins and Skellam(1954)}]{hopkins1954new}
Hopkins, B.; and Skellam, J.~G. 1954.
\newblock {A New Method for determining the Type of Distribution of Plant
  Individuals}.
\newblock \emph{Annals of Botany}.

\bibitem[{Jiang, Ren, and Ferrara(2023)}]{jiang2022retweet}
Jiang, J.; Ren, X.; and Ferrara, E. 2023.
\newblock Retweet-BERT: Political Leaning Detection Using Language Features and
  Information Diffusion on Social Networks.
\newblock \emph{ICWSM}.

\bibitem[{Jost(2017)}]{jost2017asymmetries}
Jost, J.~T. 2017.
\newblock Asymmetries abound: Ideological differences in emotion, partisanship,
  motivated reasoning, social network structure, and political trust.
\newblock \emph{Journal of Consumer Psychology}.

\bibitem[{Kariryaa et~al.(2022)Kariryaa, Rund{\'e}, Heuer, Jungherr, and
  Sch{\"o}ning}]{kariryaa2022role}
Kariryaa, A.; Rund{\'e}, S.; Heuer, H.; Jungherr, A.; and Sch{\"o}ning, J.
  2022.
\newblock The role of flag emoji in online political communication.
\newblock \emph{Social Science Computer Review}.

\bibitem[{Ke et~al.(2017)Ke, Meng, Finley, Wang, Chen, Ma, Ye, and
  Liu}]{ke2017lightgbm}
Ke, G.; Meng, Q.; Finley, T.; Wang, T.; Chen, W.; Ma, W.; Ye, Q.; and Liu,
  T.-Y. 2017.
\newblock Lightgbm: A highly efficient gradient boosting decision tree.
\newblock \emph{NeurIPS}.

\bibitem[{Kemmelmeier and Winter(2008)}]{kemmelmeier2008sowing}
Kemmelmeier, M.; and Winter, D.~G. 2008.
\newblock Sowing patriotism, but reaping nationalism? Consequences of exposure
  to the American flag.
\newblock \emph{Political Psychology}.

\bibitem[{Kerchner and Wrubel(2021)}]{riot}
Kerchner, D.; and Wrubel, L. 2021.
\newblock {U.S. Capitol Riot and \#TrumpRally Tweet IDs}.

\bibitem[{Kwak et~al.(2021)Kwak, An, Jing, and Ahn}]{kwak2021frameaxis}
Kwak, H.; An, J.; Jing, E.; and Ahn, Y.-Y. 2021.
\newblock FrameAxis: characterizing microframe bias and intensity with word
  embedding.
\newblock \emph{PeerJ Computer Science}.

\bibitem[{Lahoti, Garimella, and Gionis(2018)}]{lahoti2018joint}
Lahoti, P.; Garimella, K.; and Gionis, A. 2018.
\newblock Joint non-negative matrix factorization for learning ideological
  leaning on twitter.
\newblock In \emph{WSDM}.

\bibitem[{Lai et~al.(2022)Lai, Brown, Bisbee, Tucker, Nagler, and
  Bonneau}]{lai2022estimating}
Lai, A.; Brown, M.~A.; Bisbee, J.; Tucker, J.~A.; Nagler, J.; and Bonneau, R.
  2022.
\newblock Estimating the ideology of political youtube videos.
\newblock \emph{Political Analysis}.

\bibitem[{Li et~al.(2020)Li, Longinos, Wilson, and Magdy}]{li2020emoji}
Li, J.; Longinos, G.; Wilson, S.; and Magdy, W. 2020.
\newblock Emoji and self-identity in Twitter bios.
\newblock In \emph{NLP+CSS}.

\bibitem[{Liscio et~al.(2023)Liscio, Araque, Gatti, Constantinescu, Jonker,
  Kalimeri, and Murukannaiah}]{liscio2023does}
Liscio, E.; Araque, O.; Gatti, L.; Constantinescu, I.; Jonker, C.~M.; Kalimeri,
  K.; and Murukannaiah, P.~K. 2023.
\newblock What does a text classifier learn about morality? An explainable
  method for cross-domain comparison of moral rhetoric.
\newblock In \emph{ACL}.

\bibitem[{Liu et~al.(2023)Liu, Luo, Xu, Wei, Wei, Yu, Xiang, and
  Wang}]{liu2023ideology}
Liu, S.; Luo, Z.; Xu, M.; Wei, L.; Wei, Z.; Yu, H.; Xiang, W.; and Wang, B.
  2023.
\newblock Ideology Takes Multiple Looks: A High-Quality Dataset for
  Multifaceted Ideology Detection.
\newblock In \emph{EMNLP}.

\bibitem[{Lorena et~al.(2019)Lorena, Garcia, Lehmann, Souto, and
  Ho}]{lorena2019complex}
Lorena, A.~C.; Garcia, L.~P.; Lehmann, J.; Souto, M.~C.; and Ho, T.~K. 2019.
\newblock How complex is your classification problem? a survey on measuring
  classification complexity.
\newblock \emph{CSUR}.

\bibitem[{Maci{\`a}, Orriols-Puig, and
  Bernad{\'o}-Mansilla(2008)}]{macia2008genetic}
Maci{\`a}, N.; Orriols-Puig, A.; and Bernad{\'o}-Mansilla, E. 2008.
\newblock Genetic-based synthetic data sets for the analysis of classifiers
  behavior.
\newblock In \emph{HAIS}.

\bibitem[{McCauley and Moskalenko(2008)}]{McCauley2008}
McCauley, C.; and Moskalenko, S. 2008.
\newblock Mechanisms of Political Radicalization: Pathways Toward Terrorism.
\newblock \emph{Terrorism and Political Violence}.

\bibitem[{McPherson, Smith-Lovin, and Cook(2001)}]{mcpherson2001birds}
McPherson, M.; Smith-Lovin, L.; and Cook, J.~M. 2001.
\newblock Birds of a feather: Homophily in social networks.
\newblock \emph{Annual review of sociology}.

\bibitem[{Metaxas et~al.(2015)Metaxas, Mustafaraj, Wong, Zeng, O'Keefe, and
  Finn}]{metaxas2015retweets}
Metaxas, P.; Mustafaraj, E.; Wong, K.; Zeng, L.; O'Keefe, M.; and Finn, S.
  2015.
\newblock What do retweets indicate? Results from user survey and meta-review
  of research.
\newblock In \emph{ICWSM}.

\bibitem[{Meyer(2020)}]{meyer2020political}
Meyer, P.~H. 2020.
\newblock Political Ideology and Black-and-White Thinking.

\bibitem[{Mokhberian et~al.(2020)Mokhberian, Abeliuk, Cummings, and
  Lerman}]{mokhberian2020moral}
Mokhberian, N.; Abeliuk, A.; Cummings, P.; and Lerman, K. 2020.
\newblock Moral framing and ideological bias of news.
\newblock In \emph{SocInfo}.

\bibitem[{Newman et~al.(2021)Newman, Fletcher, Schulz, Andi, Robertson, and
  Nielsen}]{newman2021reuters}
Newman, N.; Fletcher, R.; Schulz, A.; Andi, S.; Robertson, C.~T.; and Nielsen,
  R.~K. 2021.
\newblock Reuters Institute digital news report 2021.
\newblock \emph{Reuters Institute for the Study of Journalism}.

\bibitem[{O'Hagan and Schein(2023)}]{o2023measurement}
O'Hagan, S.; and Schein, A. 2023.
\newblock Measurement in the Age of LLMs: An Application to Ideological
  Scaling.
\newblock \emph{arXiv preprint arXiv:2312.09203}.

\bibitem[{Park et~al.(2021)Park, Fisher, McGuinness, Lee, and
  McCallum}]{park2021digital}
Park, S.; Fisher, C.; McGuinness, K.; Lee, J.~Y.; and McCallum, K. 2021.
\newblock \emph{Digital news report: Australia 2021}.
\newblock News and Media Research Centre.

\bibitem[{Poole and Rosenthal(1985)}]{poole1985spatial}
Poole, K.~T.; and Rosenthal, H. 1985.
\newblock A spatial model for legislative roll call analysis.
\newblock \emph{American journal of political science}.

\bibitem[{Preo{\c{t}}iuc-Pietro et~al.(2017)Preo{\c{t}}iuc-Pietro, Liu,
  Hopkins, and Ungar}]{preoctiuc2017beyond}
Preo{\c{t}}iuc-Pietro, D.; Liu, Y.; Hopkins, D.; and Ungar, L. 2017.
\newblock Beyond binary labels: Political ideology prediction of Twitter users.
\newblock In \emph{ACL}.

\bibitem[{Radsch(2016)}]{Radsch2016}
Radsch, C. 2016.
\newblock Media Development and Countering Violent Extremism: An Uneasy
  Relationship, a Need for Dialogue.
\newblock \emph{Center for International Media Assistance}.

\bibitem[{Rao, Morstatter, and Lerman(2022)}]{rao2022partisan}
Rao, A.; Morstatter, F.; and Lerman, K. 2022.
\newblock Partisan asymmetries in exposure to misinformation.
\newblock \emph{Scientific reports}.

\bibitem[{Rao(2017)}]{rao2017red}
Rao, A.~R. 2017.
\newblock Red, blue and purple states of mind: Segmenting the political
  marketplace.
\newblock \emph{Journal of Consumer Psychology}.

\bibitem[{Rashed et~al.(2021)Rashed, Kutlu, Darwish, Elsayed, and
  Bayrak}]{rashed2021embeddings}
Rashed, A.; Kutlu, M.; Darwish, K.; Elsayed, T.; and Bayrak, C. 2021.
\newblock Embeddings-Based Clustering for Target Specific Stances: The Case of
  a Polarized Turkey.
\newblock In \emph{ICWSM}.

\bibitem[{Ravi, Vela, and Ewetz(2022)}]{ravi2022classifying}
Ravi, K.; Vela, A.~E.; and Ewetz, R. 2022.
\newblock Classifying the Ideological Orientation of User-Submitted Texts in
  Social Media.
\newblock In \emph{ICMLA}.

\bibitem[{Reiter-Haas, Kopeinik, and Lex(2021)}]{reiter2021studying}
Reiter-Haas, M.; Kopeinik, S.; and Lex, E. 2021.
\newblock Studying Moral-based Differences in the Framing of Political Tweets.
\newblock In \emph{ICWSM}.

\bibitem[{Rizoiu et~al.(2018)Rizoiu, Graham, Zhang, Zhang, Ackland, and
  Xie}]{rizoiu2018debatenight}
Rizoiu, M.-A.; Graham, T.; Zhang, R.; Zhang, Y.; Ackland, R.; and Xie, L. 2018.
\newblock \# DebateNight: The Role and Influence of Socialbots on Twitter
  During the 1st 2016 US Presidential Debate.
\newblock In \emph{ICWSM}.

\bibitem[{Samih and Darwish(2021)}]{samih2021few}
Samih, Y.; and Darwish, K. 2021.
\newblock A few topical tweets are enough for effective user stance detection.
\newblock In \emph{ACL}.

\bibitem[{Stankov(2021)}]{Stankov2021}
Stankov, L. 2021.
\newblock {From social conservatism and authoritarian populism to militant
  right-wing extremism}.
\newblock \emph{Personality and Individual Differences}.

\bibitem[{Thomas et~al.(2022)Thomas, Leggett, Kernot, Mitchell, Magsarjav, and
  Weber}]{thomas2022reclaim}
Thomas, E.~F.; Leggett, N.; Kernot, D.; Mitchell, L.; Magsarjav, S.; and Weber,
  N. 2022.
\newblock Reclaim the Beach: How Offline Events Shape Online Interactions and
  Networks Amongst Those Who Support and Oppose Right-Wing Protest.
\newblock \emph{Studies in Conflict \& Terrorism}.

\bibitem[{Tomkins(1963)}]{tomkins1963left}
Tomkins, S. 1963.
\newblock Left and right: A basic dimension of ideology and personality.

\bibitem[{Van~der Vegt et~al.(2021)Van~der Vegt, Mozes, Kleinberg, and
  Gill}]{van2021grievance}
Van~der Vegt, I.; Mozes, M.; Kleinberg, B.; and Gill, P. 2021.
\newblock The grievance dictionary: Understanding threatening language use.
\newblock \emph{Behavior research methods}.

\bibitem[{van Vliet, T{\"o}rnberg, and Uitermark(2020)}]{van2020twitter}
van Vliet, L.; T{\"o}rnberg, P.; and Uitermark, J. 2020.
\newblock The Twitter parliamentarian database: Analyzing Twitter politics
  across 26 countries.
\newblock \emph{PLoS one}.

\bibitem[{Van~Vliet, T{\"o}rnberg, and Uitermark(2021)}]{van2021political}
Van~Vliet, L.; T{\"o}rnberg, P.; and Uitermark, J. 2021.
\newblock Political Systems and Political Networks: The Structure of
  Parliamentarians’ Retweet Networks in 19 Countries.
\newblock \emph{International Journal of Communication}.

\bibitem[{Wang et~al.(2021)Wang, Wu, Weimer, and Zhu}]{wang2021flaml}
Wang, C.; Wu, Q.; Weimer, M.; and Zhu, E. 2021.
\newblock FLAML: A fast and lightweight automl library.
\newblock \emph{MLSys}.

\bibitem[{Wang and Inbar(2021)}]{wang2021moral}
Wang, S.-Y.~N.; and Inbar, Y. 2021.
\newblock Moral-language use by US political elites.
\newblock \emph{Psychological Science}.

\bibitem[{Wikipedia(2023)}]{Q+Ashow2023}
Wikipedia. 2023.
\newblock {Q+A (Australian talk show)}.
\newblock \url{https://en.wikipedia.org/wiki/Q\%2BA_(Australian_talk_show)}.
\newblock Accessed: 2023-04-27.

\bibitem[{Xi et~al.(2020)Xi, Ma, Liou, Steinert-Threlkeld, Anastasopoulos, and
  Joo}]{xi2020understanding}
Xi, N.; Ma, D.; Liou, M.; Steinert-Threlkeld, Z.~C.; Anastasopoulos, J.; and
  Joo, J. 2020.
\newblock Understanding the political ideology of legislators from social media
  images.
\newblock In \emph{ICWSM}.

\bibitem[{Xiao et~al.(2020)Xiao, Song, Xu, Ren, and Sun}]{xiao2020timme}
Xiao, Z.; Song, W.; Xu, H.; Ren, Z.; and Sun, Y. 2020.
\newblock TIMME: Twitter ideology-detection via multi-task multi-relational
  embedding.
\newblock In \emph{KDD}.

\bibitem[{Zandt(2022)}]{mbfc}
Zandt, D. 2022.
\newblock Media Bias/Fact Check.
\newblock \url{https://mediabiasfactcheck.com/about}.
\newblock Accessed: 2022-04-08.

\end{thebibliography}
\end{document}